\documentclass[useAMS,usenatbib]{mn2e}
\usepackage{amsmath}
\usepackage{amsfonts}
\usepackage{graphics}
\usepackage{subfigure}
\usepackage{epsfig}
\usepackage{rotate}
\usepackage{ifpdf}
\usepackage{graphicx,amssymb}

\title[Regular and chaotic orbits in barred galaxies - I.]
{Regular and chaotic orbits in barred galaxies - I. Applying the SALI/GALI method to explore their distribution in several models}

\author[Manos $\&$ Athanassoula]{T.~Manos$^{\mathrm{a},\mathrm{b},\mathrm{c}}\thanks{E-mail:thanosm@master.math.upatras.gr (TM); lia@oamp.fr (EA)}$ and
E.~Athanassoula$^{\mathrm{a}}$\\
$^a$ LAM, UMR6110, CNRS/Universit\'e de Provence, 38 rue
Joliot Curie, 13388 Marseille C\'edex 13, France.\\
$^b$ Center for Research and Applications of Nonlinear Systems, Department of Mathematics, University of Patras,
GR--26500, Patras, Greece.\\
$^c$ University of Nova Gorica, School of applied sciences, Vipavska 11c, SI-5270, Ajdov\v{s}\v{c}ina, Slovenia.\\}
\date{Released 2011 March 16}
\pagerange{1--16} \pubyear{2011}

\def\LaTeX{L\kern-.36em\raise.3ex\hbox{a}\kern-.15em
    T\kern-.1667em\lower.7ex\hbox{E}\kern-.125emX}

\begin{document}
\maketitle

\begin{abstract}
The distinction between chaotic and regular behavior of orbits in galactic
models is an important issue and can help our understanding of galactic
dynamical evolution. In this paper, we deal with this issue by applying the
techniques of the Smaller (and Generalized) ALingment Indices, SALI (and GALI),
to extensive samples of orbits obtained by integrating numerically the
equations of motion in a barred galaxy potential. We estimate first the
fraction of chaotic and regular orbits for the two--degree--of--freedom (DOF)
case (where the galaxy extends only in the $(x,y)$--space) and show that it is
a non--monotonic function of the energy. For the three DOF extension of this
model (in the $z$--direction), we give similar estimates, both by exploring
different sets of initial conditions and by varying the model parameters, like
the mass, size and pattern speed of the bar. We find that regular motion is more
abundant at small radial distances from the center of the galaxy, where the
relative non-axisymmetric forcing is relatively weak, and at small distances
from the equatorial plane, where trapping around the stable periodic
orbits is important. We also find that the variation of the bar pattern
speed, within a realistic range of values, does not affect much the phase
space's fraction of regular and chaotic motions. Using different sets of initial
conditions, we show that chaotic motion is dominant in galaxy models whose bar
component is more massive, while models with a fatter or thicker bar present
generally more regular behavior. Finally, we find that the fraction of orbits
that are chaotic correlates strongly with the bar strength.
\end{abstract}


\section{Introduction} \label{intro}

Exploring the nature of orbits in galaxies constitutes a very important issue,
not only because of the evident astronomical interest in classifying the types
of orbits that exist in such systems, but also because orbits are needed for
constructing self-consistent models of galaxies. In order to study and
understand the structure and dynamics of a galaxy it is necessary to identify
the type of dynamics characterizing the motion of stars (regular or chaotic)
and estimate the percentages of each type in the phase space of the galaxy.

From previous works that describe galaxies and their star motion, it is
well-known that the analysis of periodic orbits, and their stability, can
provide very useful information about galaxy structure. Stable periodic orbits
are associated with regular motion, since they are surrounded by tori of
quasi-periodic motion. Thus regular orbits are trapped in the vicinity of the
parent periodic orbit. On the other hand unstable periodic orbits breed chaos.
If we pick an orbit in the immediate phase-space vicinity of a chaotic one, we
find that the two orbits will diverge exponentially with time. Such chaotic
orbits fill up all the phase space region that is available to them.
Nevertheless, recent results in galactic dynamics show that there are chaotic
orbits that can support galaxy features, even thin features like the outer
parts of bars, spirals and rings \citep{KauCo, PAQ1997, Pats:2006, Rom:2006,
Rom:2007, VKS2, VTE, VHC, ConHa, Athan:2009a, Athan:2009b, Athan:2010,
HaKa:2009, Tsou:2009}. These orbits are often called ``sticky" and their true
chaotic nature takes very long to be revealed \citep{Con_spr}.
\cite{Athan:2010} termed these orbits ``confined chaotic orbits" because the
structures they generate are well confined in configuration space. There are
also several recent related results showing that strong local instability does
not mean diffusion in phase space (e.g. \citealt{Gio:2004, CinGiochap:2008,
Cachuchoetal:2010}). In our next paper, paper II, we will focus on this special
category of chaotic orbits discussing their significance from an observational
point of view.

Ferrers' potentials \citep{Fer} have proven very efficient for studying the
main properties of real bars. The detection of periodic orbits and their
stability in these models have been studied in detail by many researchers (e.g.
\citealt{ABMP,Pfe:1,SPA02a,SPA02b,PSA02,PSA03a,PSA03b}). It is well-known
that stable periodic orbits of low period possess in their neighborhood
sizeable domains of quasiperiodic (or regular) motion. A number of questions
arises, therefore, concerning these regions of stability: How far from the
stable periodic orbit, can regular motion be sustained? Where are chaotic
regions located in phase space and configuration space? What are the model's
parameters that favor large islands of stability around the main stable
periodic orbits? In this paper we plan to address such questions and to provide
some tentative answers.

Chaos arises in many areas of dynamical systems and its effect is generally
related to the type of nonlinear terms present in the model's equations of
motion (see \citealt{Con_spr} for a review). The dynamics of stars in galactic
potentials is a particular case, where nonlinear terms play an important role.
The distinction between regular and chaotic motion is not trivial
and becomes more subtle in systems of many DOF. Thus, it is necessary to use
fast and precise methods to identify the nature of orbits in such models.

Many methods have been developed over the years dealing with this problem. The
inspection of successive intersections of an orbit with a Poincar\'{e} surface
of section (PSS) \citep{LL} has been particularly useful for two DOF
Hamiltonian systems. One of the most popular methods of chaos detection is the
computation of the maximal ``Lyapunov Characteristic Exponent" (LCE)
\citep{Ose,Ben:1980a,Ben:1980b,Sko:LE}. Along the same lines as LCE, several
other methods of chaos detection have been proposed based on the study of the
behavior of deviation vectors: We may mention, e.g. the ``Fast Lyapunov
Indicator" \citep{Fro:3,Fro:4} and the ``Mean Exponential Growth of Nearby
Orbits" \citep{Cin:1,Cin:2}. On the other hand, there also exist methods based
on the analysis of time series constructed by the coordinates of each orbit
such as the ``Frequency Map Analysis" \citep{Las:1,Las:2,Las:3}. More details
about these and other relative methods can be found in the book of
\cite{Con_spr} and in the review of \cite{Sko:LE} .

In the present paper, we use a method called the ``Smaller ALingment Index"
(SALI), based on the properties of two deviation vectors of an orbit for the
quick and efficient distinction of chaotic motion, originally introduced by
\cite{sk:1}. It has been successfully applied to different dynamical systems
\citep{sk:1,sk:2,VKS1,sk:3,sk:4,KVC,sk:5,Pan:1,SESS,Ant:2,Ant:3,CDLMV,VKS2,VHC,KEV,Man:4,Man:6,Stra2009,Macek2010}.
In every case, it was confirmed to be a fast and reliable indicator of the
chaotic or ordered nature of orbits. We also use SALI's recent generalization,
the so-called ``Generalized ALignment Index" (GALI) introduced by \cite{sk:6},
using a set of $p$ initially linearly independent deviation vectors of the
system. Thus, following more deviation vectors, we manage to acquire more extra
information about the complexity of the regular motion, i.e. the dimensionality
of the invariant torus (or simply torus from now on) on which the orbit lies
\citep{SkoBouAnto,Man:5,Man:7,BouManChris,ManRuf}.

The paper is organized as follows: Sections~\ref{Lyap_exp} and \ref{SALIGALI}
are devoted to the chaos detectors we use, i.e. the Lyapunov spectra and the
SALI/GALI methods. In Section~\ref{Model_gal}, we present in detail the model,
which is composed by a bulge, a disk and a Ferrers bar \citep{Fer}. In
Section~\ref{2dof}, we present our results for the two DOF restriction of
the general model, calculating the regular and chaotic orbits and constructing
also charts of the different types of motion in the associated phase space.
Section~\ref{3dof} is dedicated to the study of the full three DOF model in
terms of computing percentages of regular and chaotic orbits, selecting
different sets of initial conditions and varying the parameters of the bar
component. Thus, with the help of the SALI method, we first distinguish the
true nature of the studied orbits in phase space (Section~\ref{3Dphase_space})
and correlate the bar force with the presence of larger
amount of chaotic motion in phase space. Finally, in Section~\ref{Conclusions},
we summarize our conclusions.

\section{Lyapunov Exponents}
\label{Lyap_exp}

The Lyapunov Characteristic Exponents (LCEs) or Lyapunov Characteristic Numbers
(LCN) or more simple Characteristic Exponents are very important for the study
of dynamical systems, for distinguishing between regular and chaotic behavior
of orbits in phase space \citep{Ben:1980a,Ben:1980b,LL,Pett,Sko:LE}. In
practice, the LCEs describe the rate of separation of infinitesimally close
trajectories. The mathematical definition of LCE relies on Oselede\v{c}'s
multiplicative theorem \citep{Ose}.

A flow $\mathbf{x}(t)$ generated by an autonomous first-order system is given
by:
\begin{equation}\label{dif_eq}
\frac{d\mathbf{x}(t)}{dt}=F(\mathbf{x}(t)),
\end{equation}
where $F$ is its velocity field. Let us consider a trajectory in
$M$--dimensional phase space together with a nearby trajectory, with initial
conditions $\mathbf{x_{0}}$ and $\mathbf{x_{0}}+\Delta \mathbf{x_{0}}$,
respectively. These evolve with time yielding the tangent vector $\Delta
\mathbf{x}(\mathbf{x_{0}},t)$ with its Euclidean norm:
\begin{equation}\label{euc_norm}
    d(\mathbf{x_{0}})=\| \Delta \mathbf{x}(\mathbf{x_{0}},t)\|.
\end{equation}
Writing $\Delta \mathbf{x}=(\Delta x_{1},...,\Delta x_{M})\equiv
\mathbf{w}$, the time evolution for $\mathbf{w}$ if found by linearizing
(\ref{dif_eq}), to obtain the variational equations:
\begin{equation}\label{var_eq}
\frac{d\mathbf{w}}{dt}=J(\mathbf{x}(t))\mathbf{w},
\end{equation}
where $J(\mathbf{x}(t))=\partial F / \partial \mathbf{x}$ is the Jacobian
matrix of the $F(\mathbf{x})$. The mean exponential rate of divergence of two
initially close trajectories is:
\begin{equation}\label{LE}
\sigma (\mathbf {x}_{0},w)=\lim_{t\rightarrow \infty}(\frac{1}{t}) \ln \frac{d(\mathbf{x}_{0},t)}{d(\mathbf{x}_{0},0)}.
\end{equation}
It can be shown that $\sigma$ exists and is finite. Furthermore, there is an
$M$--dimensional basis ${\hat{e}_i}$ of $\mathbf{w}$ such that for any
$\mathbf{w}$, $\sigma$ takes one of the $M$ (possibly non-distinct) values
$\sigma_{i}(\mathbf{x}_{0})=\sigma(\mathbf{x}_{0},\hat{e}_{i}), \quad \forall
i=1,2,...,M,$ which are the Lyapunov characteristic exponents. These can be
ordered by size $\sigma_{1} \geq \sigma_{2} ... \geq \sigma_{M}$.

\section{The method of the SALI (and GALI) spectra}
\label{SALIGALI}

Let us consider a Hamiltonian flow of $N$ DOF, an orbit in the
$2N$--dimensional phase space with initial condition
$P(0)=(x_{1}(0),x_{2}(0),...,x_{2N}(0))$ and two deviation vectors
$\mathbf{w}_{1}(0)$, $\mathbf{w}_{2}(0)$ from the initial point $P(0)$. In
order to compute the SALI for that orbit one has to follow the time evolution
of the orbit itself as well as the two deviation vectors
$\mathbf{w}_{1}(t),\mathbf{w}_{2}(t)$ which initially point in two arbitrary
directions. The evolution of the deviation vectors is given by the variational
equations (\ref{var_eq}) of the flow. At every time step the two deviation
vectors $\mathbf{w}_{1}(t)$ and $\mathbf{w}_{2}(t)$ are normalized by setting:
\begin{equation}\label{norm_dev}
\hat{\textbf{w}}_{i}(t)=\frac{\mathbf{w}_{i}(t)}{\|\mathbf{w}_{i}(t)\|}, \quad i=1,2
\end{equation}
and the SALI is computed as \citep{sk:1}:
\begin{equation}\label{eq:SALI}
    \text{SALI}(t)=min
    \left\{\left\|\hat{\textbf{w}}_{1}(t)+\hat{\textbf{w}}_{2}(t)\right\|,\left\|\hat{\textbf{w}}_{1}(t)-\hat{\textbf{w}}_{2}(t)\right\|\right\}.
\end{equation}
The properties of the time evolution of the SALI rapidly distinguish between
ordered and chaotic motion as follows: The SALI fluctuates around a non-zero
value for ordered orbits, while it tends exponentially to zero for chaotic
orbits \citep{sk:3,sk:4}. In general, two different initial deviation vectors
become tangent to different directions on the torus, producing different
sequences of vectors, so that the SALI always fluctuates around positive
values. On the other hand, for chaotic orbits, any two initially different
deviation vectors in time tend to \textit{align} in the direction defined by
the maximal LCE (mLCE). Hence, they either coincide with each other, or become
opposite, which leads to the SALI falling exponentially to zero. Thus, this
completely different behavior of the SALI helps us distinguish between ordered
and chaotic motion in Hamiltonian systems of any dimensionality. An analytical
study of SALI's behavior for such orbits was carried out in \cite{sk:5}, where
it was shown that SALI $\varpropto e^{-(\sigma_{1}-\sigma_{2})t}$,
$\sigma_{1},\sigma_{2}$ being the two largest Lyapunov exponents.

The usual technique to decide whether an orbit can be called chaotic or regular
is to check, after some time interval, if its SALI has become less than a very
small threshold value. In the following we will take this value to be equal to
$10^{-8}$. Depending on the location of the orbit, this limit can be reached
more or less fast, as there are phenomena that can hold off the final
characterization of the orbit, and certain orbits behave as regular for long
times before finally drifting away from regular regions and starting to wander
in a chaotic domain.

The generalized alignment index of order $p$ (GALI$_p$) is determined through
the evolution of $2 \leq p \leq 2N$ initially linearly independent deviation
vectors $\textbf{w}_i(0), i = 1,2,...,p$, so it is related to the computation
of many LCEs rather than just the maximal one. The evolved deviation vectors
$\textbf{w}_i(t)$ are normalized every few time steps in order to avoid
overflow problems, but their directions are left intact. Then, according to
\cite{sk:6}, GALI$_p$ is defined to be the volume of the $p$–-parallelogram
having as edges the $p$ unitary deviation vectors $\hat{\textbf{w}}_i(t), i =
1,2,...,p$:
\begin{equation}\label{GALI:0}
    \text{GALI}_{p}(t)=\parallel \hat{\textbf{w}}_{1}(t)\wedge \hat{\textbf{w}}_{2}(t)
    \wedge ... \wedge \hat{\textbf{w}}_{p}(t) \parallel.
\end{equation}
From the definition of GALI$_p$ it becomes evident that if at least two of the
deviation vectors become linearly dependent, the wedge product in
Eq.~(\ref{GALI:0}) becomes zero and the GALI$_p$ vanishes.

In the case of a chaotic orbit, all deviation vectors tend to become
\textit{linearly dependent}, aligning in the direction defined by the maximal
Lyapunov exponent and GALI$_{p}$ tends exponentially to zero following the law:
\begin{equation}\label{GALI:1}
\text{GALI}_{p}(t) \sim e^{-[(\sigma_{1}-\sigma_{2})+(\sigma_{1}-\sigma_{3})+...+(\sigma_{1}-\sigma_{p})]t},
\end{equation}
where $\sigma_1> \ldots >\sigma_p$ are approximations of the first $p$ largest
Lyapunov exponents. In the case of regular motion, on the other
hand, all deviation vectors tend to fall on the $N$--dimensional tangent space
of the torus, where the motion is quasiperiodic. Thus, if we start with $p\leq
N$ general deviation vectors, these will remain \textit{linearly independent}
on the $N$--dimensional tangent space of the torus, since there is no
particular reason for them to become aligned. As a consequence, GALI$_{p}$ in
this case remains practically constant for $p\leq N$. On the other hand, for
$p>N$, GALI$_{p}$ tends to zero, since some deviation vectors will eventually
become \textit{linearly dependent}, following power laws that depend on the
dimensionality of the torus. According to \cite{ChrisBou} and
\cite{SkoBouAnto} one
obtains the following formula for the GALI$_{p}$, associated with quasiperiodic
orbits lying on $k$--dimensional tori (where $k$ is potentially equal to, or
smaller than the system's size):
\begin{equation}\label{GALI:2}
   \text{GALI}_{p}(t) \sim \left\{
                         \begin{array}{ll}
                            \text{constant}, & \hbox{if $2\leq p\leq k$}\\
                           \frac{1}{t^{p-k}}, & \hbox{if $k<p\leq 2N-k$} \\
                           \frac{1}{t^{2(p-N)}}, & \hbox{if $2N-k<p\leq 2N$}.
                         \end{array}
                       \right.
\end{equation}
In the case  where $k=N$, GALI$_p$ remains constant for $2\leq p\leq N$ and
decreases to zero as  $\sim 1/t^{2(p-N)}$ for $N < p \leq 2N $. An efficient
way to calculate GALI$_{p}$ is by multiplying the singular values $z_i,
i=1,...,p$, computed through a Singular Value Decomposition procedure of the
matrix formed by the deviation vectors $\hat{\textbf{w}}_{i}, i=1,...,p$
\citep{LDI,SkoBouAnto}:
\begin{equation}\label{svd_calc}
    \text{GALI}_{p}=\prod_{i=1}^{p} z_i.
\end{equation}
The method has been applied successfully in several Hamiltonian systems like
the FPU lattice \citep{SkoBouAnto} and coupled symplectic maps
\citep{BouManChris} for the detection not only of regular and chaotic motion
but also the dimensionality of the torus on which a regular trajectory lies on.

Practically, the SALI is equivalent to GALI$_2$ and the distinction between
regular and chaotic motion in the 2D Ferrers barred model (two DOF) can be done
with either one of them. For the full 3D version of the model, we also use
GALI$_3$, depending on the properties of the model (if its phase space is
dominated by large chaotic or regular areas) and on our goals. GALI$_3$
generally demands more CPU time, since it is necessary to follow the evolution
of three deviation vectors instead of two. For the case of chaotic trajectories,
however, GALI$_3$ decays much faster and the calculation can stop well before
the end of total time interval. Summarizing therefore, if someone had to
estimate the amount of chaotic vs. regular regions in phase space, GALI$_3$
would be more efficient for models where chaos is dominant, while SALI would be
preferable for models with large regions of order. In our runs, we have used
both SALI (GALI$_2$) for the general description of chaotic vs. ordered regions,
and GALI$_3$ to follow specific regular orbits and study the dimensionality of
the torus on which they lie.

\section{The model potential}
\label{Model_gal}

The motion of a test particle in a 3D rotating model of a barred galaxy is
governed by the Hamiltonian:
\begin{equation}\label{eq:Hamilton}
   H=\frac{1}{2} (p_{x}^{2}+p_{y}^{2}+p_{z}^{2})+ V(x,y,z) -
   \Omega_{b} (xp_{y}-yp_{x}).
\end{equation}
The bar rotates around its $z$--axis (short axis), while the $x$--direction is
along the major axis and the $y$ along the intermediate axis of the bar. The
$p_{x},p_{y}$ and $p_{z}$ are the canonically conjugate momenta, $V$ is the
potential, $\Omega_{b}$ represents the pattern speed of the bar and $H$ is the
total energy of the orbit in the rotating frame of reference (Jacobi constant).
The corresponding equations of motion are:
\begin{flalign}\label{eq_motion}
 \dot{x}& =  p_{x} + \Omega_{b} y,& \quad  \dot{y}& =  p_{y} - \Omega_{b}x,&    \dot{z}& = p_{z},&\\
 \dot{p_{x}}& =  -\frac{\partial V}{\partial x} + \Omega_{b} p_{y},&  \dot{p_{y}}& = -\frac{\partial V}{\partial y} -
 \Omega_{b} p_{x},&  \dot{p_{z}}& = -\frac{\partial V}{\partial z}.& \nonumber
\end{flalign}
The potential $V$ of our model consists of three components:
\begin{enumerate}
 \item A disc, represented by a Miyamoto-Nagai disc \citep{Miy.1975}:
 \begin{equation}\label{Miy_disc}
  V_D=- \frac{GM_{D}}{\sqrt{x^{2}+y^{2}+(A+\sqrt{z^{2}+B^{2}})^{2}}},
\end{equation}
where $M_{D}$ is the total mass of the disc, $A$ and $B$ are its horizontal
and vertical scale-lengths, and $G$ is the gravitational constant.
\item A bulge, which is modeled by a Plummer sphere \citep{Plum} whose
    potential is:
\begin{equation}\label{Plum_sphere}
    V_S=-\frac{G M_{S}}{\sqrt{x^{2}+y^{2}+z^{2}+\epsilon_{s}^{2}}},
\end{equation}
where $\epsilon_{s}$ is the scale-length of the bulge and $M_{S}$ is its
total mass.
\item A triaxial Ferrers bar \citep{Fer}, the density $\rho(x)$ of which
    is:
\begin{equation}\label{Ferr_bar}
  \rho(x)=\begin{cases}\rho_{c}(1-m^{2})^{2} &, m<1  \\
              \qquad 0 &, m\geq1 \end{cases},
\end{equation}
where $\rho_{c}=\frac{105}{32\pi}\frac{G M_{B}}{abc}$ is the central density, $M_{B}$ is the total mass of the bar and
\begin{equation}\label{Ferr_m}
  m^{2}=\frac{x^{2}}{a^{2}}+\frac{y^{2}}{b^{2}}+\frac{z^{2}}{c^{2}},
\qquad a>b>c> 0,
\end{equation}
with $a,b$ and $c$ being the semi-axes. The corresponding potential is:
\begin{equation}\label{Ferr_pot}
    V_{B}= -\pi Gabc \frac{\rho_{c}}{n+1}\int_{\lambda}^{\infty}
    \frac{du}{\Delta (u)} (1-m^{2}(u))^{n+1},
\end{equation}
where
\begin{equation}\label{mu2}
m^{2}(u)=\frac{x^{2}}{a^{2}+u}+\frac{y^{2}}{b^{2}+u}+\frac{z^{2}}{c^{2}+u},
\end{equation}
\begin{equation}\label{Delta}
\Delta^{2} (u)=({a^{2}+u})({b^{2}+u})({c^{2}+u}),
\end{equation}
$n$ is a positive integer (with $n=2$ for our model) and $\lambda$ is the
unique positive solution of:
\begin{equation}\label{mu2_lamda}
    m^{2}(\lambda)=1,
\end{equation}
outside of the bar ($m \geq 1$), and $\lambda=0$ inside the bar. The
corresponding forces are given analytically by \cite{Pfe:1}.
\end{enumerate}
This model has been used extensively for orbital studies by \cite{Pfe:1},
\cite{SPA02a,SPA02b} and by \cite{PSA02,PSA03a,PSA03b}. Our so--called
\textit{standard} (S) model has the following values of parameters\ $G$=1,
$\Omega_{b}$=0.054 (54 $km\cdot sec^{-1} \cdot kpc^{-1}$), $a$=6, $b=$1.5,
$c$=0.6, $A$=3, $B$=1, $\epsilon_{s}$=0.4, $M_{B}$=0.1, $M_{S}$=0.08,
$M_{D}$=0.82, both for its two DOF and three DOF versions. The units used are:
1 $kpc$ (length), 1000 $km\cdot sec^{-1}$ (velocity), 1 $Myr$ (time), $2 \times
10^{11} M_{\bigodot}$ (mass). The total mass $G(M_{S}+M_{D}+M_{B})$ is set
equal to 1.

Our study is mainly focused on understanding the effect of the bar on the
dynamics of the model's phase space, by varying the bar mass $M_B$, its
semiaxis lengths $b,c$ and its pattern speed $\Omega_{b}$. In order to do this
we have considered three more models: models $C$, $B$ and $M$ whose short
$z$--semiaxis ($c$--parameter), intermediate $y$--semiaxis ($b$--parameter) and
mass of the bar ($M_{B}$--parameter) are twice those of the initial reference
model. For each of these four ($S,C,B$ and $M$) models we launch three
different sets of initial conditions (distributions $I,II,III$). Hence, we
consider the three \textit{standard} models ($IS,IIS,IIIS$) and their
variations ($IC,IIC,IIIC$), ($IB,IIB,IIIB$) and ($IM,IIM,IIIM$), depending on
the varying parameter. These sets of initial conditions will be discussed in
detail in Section~\ref{3Dphase_space}. The maximal time of integration of the
orbits through the equations of motion and the variational equations is set to
be T=10,000$Myr$ (10 billion yrs), that corresponds to a time less than, but of
the order of one Hubble time. We chose to take a maximum integration time that
is the same for all orbits, independent of their energy. As a result, some of
the orbits -- and particularly the ones in the innermost regions of the galaxy
-- may have been calculated for a larger multiple of their own dynamical times
than others. This choice, however, is necessary because it allows comparison
with, or applications to a galaxy or a simulation, where all information refers
to one time only, the same for all orbits. It was therefore also adopted in
previous orbital structure studies in galactic models (e.g.
\citealt{ZS1,ZS2,Val:2010}). This choice is essential in the case of some
sticky orbits which may appear regular if calculated till e.g. 10,000$Myr$, but
show clear signs of chaoticity when calculated for much longer times, i.e.
orbits whose diffusion time scale is equal to many Hubble times. Such times,
however, are irrelevant for our applications, and thus these orbits should not
be counted as chaotic, even if in the long run they may become so. These issues
will be discussed further in Paper II.

\begin{figure}\centering
  \includegraphics[height=0.275\textheight]{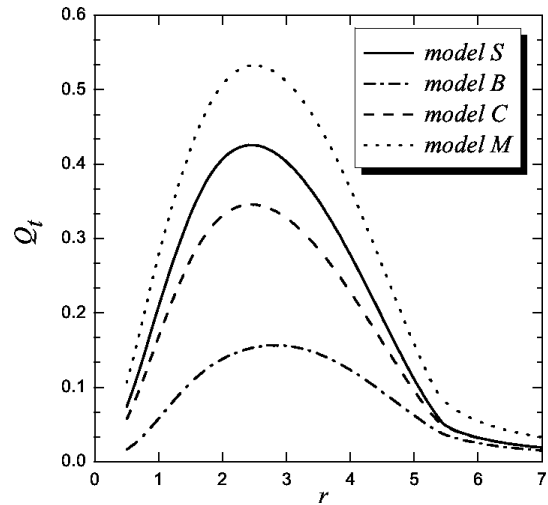}
  \caption{The quantity $Q_t$ as a function of radius. Its maximum value corresponds to
  $Q_b$, which gives an estimate of the relative strength of the
    non-axisymmetric forces. We see that the increase
  of the mass of the bar component (model $M$: dotted line) corresponds to ``stronger bar'' compared to the
  initial standard model (model $S$: solid line). On the other hand, when we increase the short $z$--semiaxis
  ($c$--parameter - model $C$: dashed line), or the intermediate $y$--semiaxis ($b$--parameter model $B$: dash-dotted line)
  the bar becomes weaker.}
  \label{barstre}
\end{figure}

The relative strength of the non-axisymmetric forces can be estimated by the
quantity $Q_{t}$, defined as:
\begin{equation}\label{Qt}
  Q_t(r) =(\partial \Phi(r,\theta)/\partial \theta )_{max}/(r \partial \Phi_0/\partial r),
\end{equation}
The maximum of $Q_t$ over all radii shorter than the bar extent is often
referred to as $Q_b$, and used as a measure of the bar strength
\citep[e.g.][]{ButaBK03, ButaLS04, LaurikainenSB04,ButaVSL05,DurbalaBSVM09}.
From now on, we will for simplicity and conciseness, often refer to the ``strong
non-axisymmetric forcings'' simply as ``strong bars''. In Fig.~\ref{barstre} we
show the above quantity for the 4 models we are going to study and discuss in
this paper. From the quantity $Q_{t}$ it becomes clear that the increase of the
mass of the bar component (model $M$: dotted line) gives rise to a ``stronger
bar'' compared to the initial standard model (model $S$: solid line). On the
other hand, an increase of the short $z$--semiaxis ($c$--parameter - model $C$:
dashed line) or intermediate $y$--semiaxis ($b$--parameter - model $B$:
dot-dashed line) has as a result ``weaker bars". Hence, how do ``strong bars"
affect the relative regular and chaotic motion in the phase space of the full
three DOF model? Before answering this question, let us first study and
comprehend the dynamics of the two DOF model.
\section{The two DOF model potential}
\label{2dof}

\begin{figure*}\centering
\includegraphics[height=0.225\textheight]{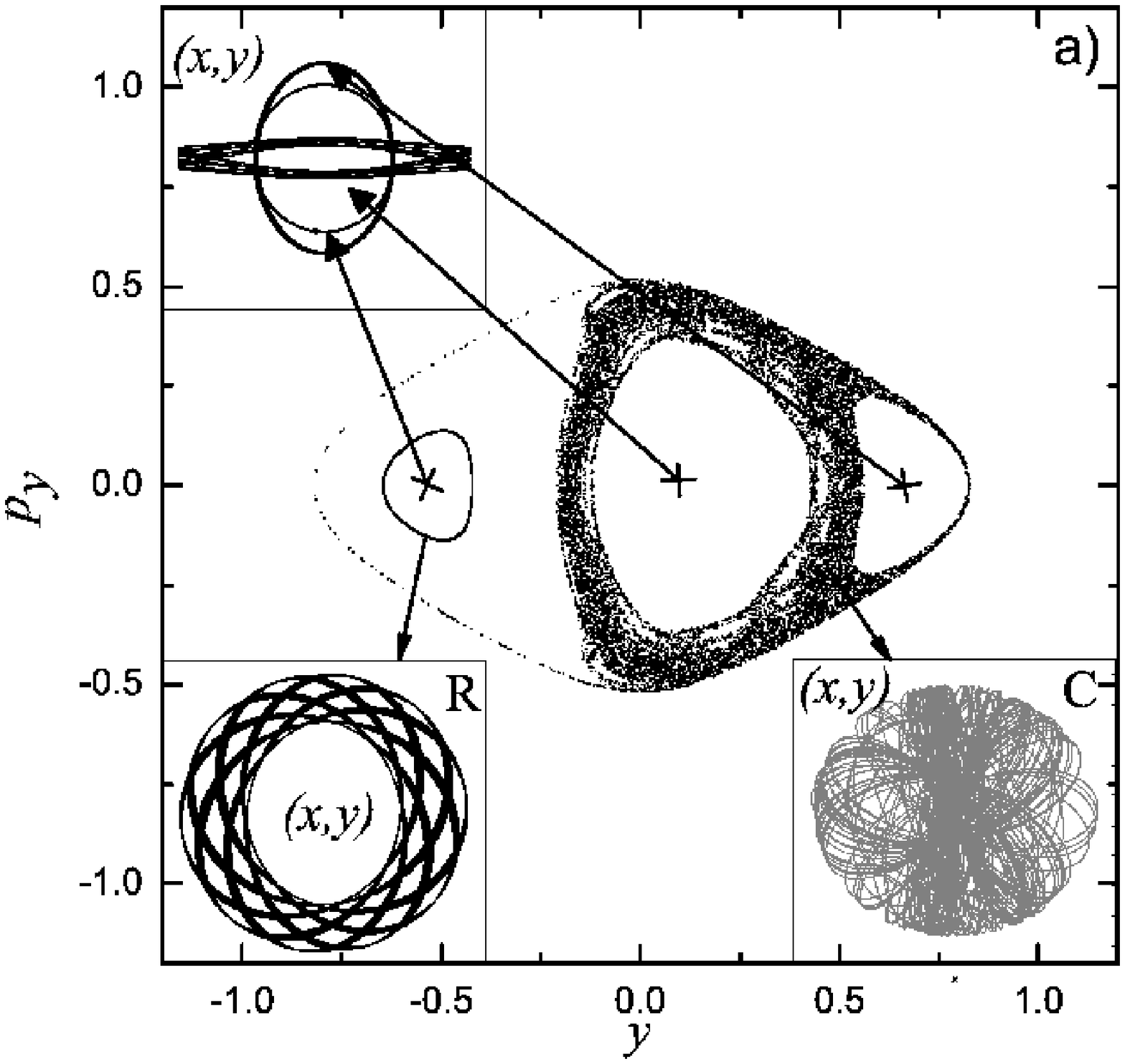}
\includegraphics[height=0.225\textheight]{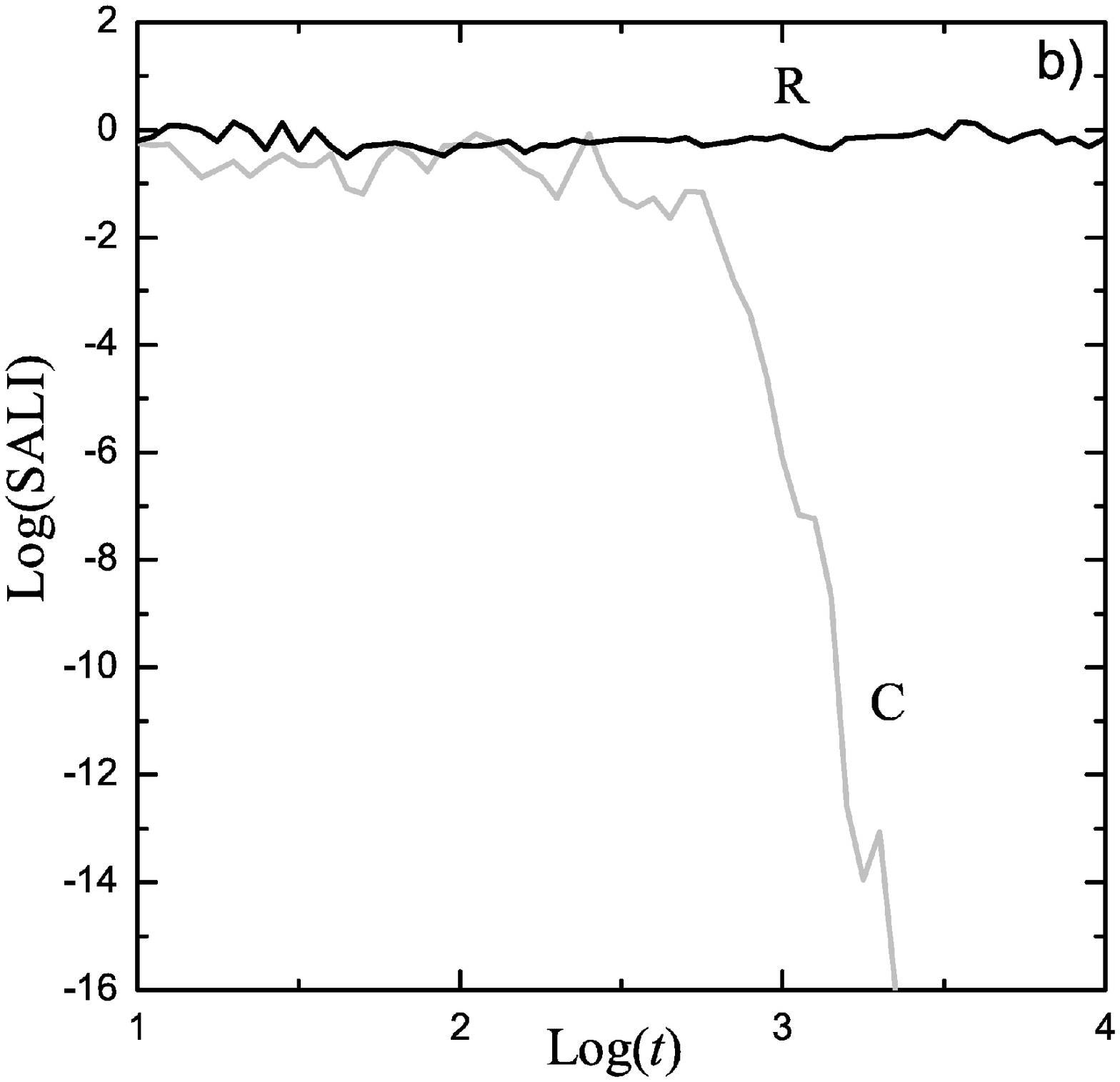}
\includegraphics[height=0.225\textheight]{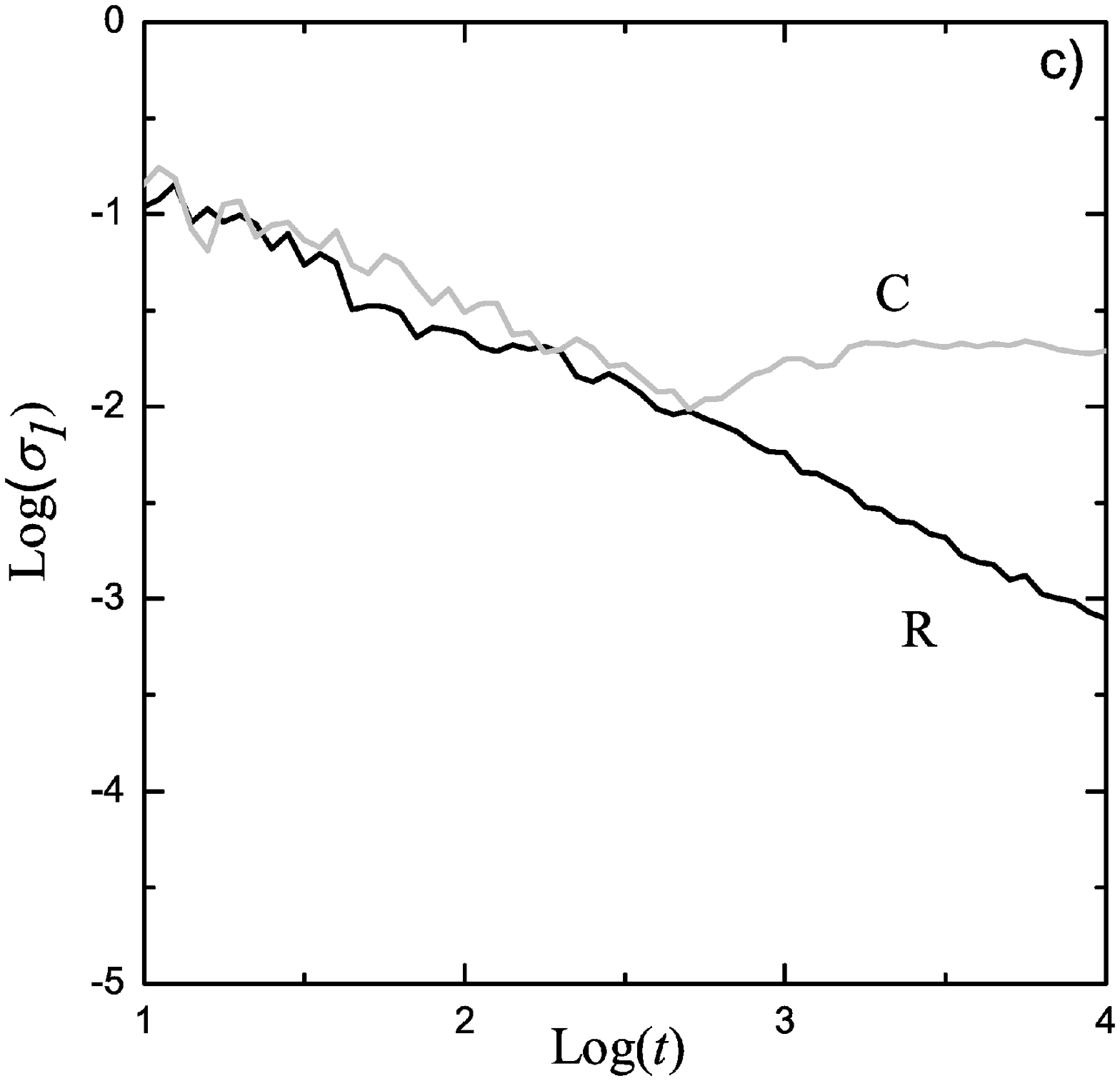}
\caption{Analysis of orbits in the two DOF case: a) Poincar\'{e} surface of
section in the $(y,p_{y})$--plane  for a regular (R--black color) and a chaotic
(C--gray color) orbit ($H=-0.36$). The points of the successive intersections
of the regular orbit R with this plane create a closed curve, while the points
of the chaotic orbit C fill with scattered points all its available region of
motion. The orbits can be seen in the two inset figures in the bottom of the
panel. In the top inset figure, we show examples of the three different
morphologies of the regular orbits around the three main periodic orbits in the
centers of the three main islands of stability. Note that the basic barred
shape in the $(x,y)$--plane is mainly provided by trajectories from the central
stable region, around the family of periodic orbits x1 (orbits elongated along
the bar's major axis). The other two stability islands give either
elliptical-like orbits elongated perpendicular to the bar orbits
(right island - family x2), or orbits which are elliptical-like  but
retrograde with respect the bar's pattern speed
(left island - family x4). b) The corresponding evolution of the SALI for the R
and C orbits of panel a. For the chaotic one C (gray line), the SALI decays
exponentially fast to zero, while for the ordered one R (black line) it
fluctuates around a non-zero number. c) The corresponding evolution of the
maximal Lyapunov exponents $\sigma_1$ for the orbits R and C. For the
chaotic orbit C (gray line) this tends after some transient time to a
non-zero value, while for the
regular orbit R (black color) it tends linearly to zero. Note that the
$\text{Log(SALI)}$ in $y$--axis of panel b ranges from -16 to 2, while the
$\text{Log}(\sigma_1)$ in panel c ranges from -5 to 0.} \label{pss_SALI_LE}
\end{figure*}

The two DOF Ferrers model is described by the Hamiltonian Eq.~
(\ref{eq:Hamilton}) setting $(z,p_z)=(0,0)$, i.e.:
\begin{equation}\label{eq:2dof}
   H=\frac{1}{2} (p_{x}^{2}+p_{y}^{2})+ V(x,y) - \Omega_{b} (xp_{y}-yp_{x}).
\end{equation}
In the two bottom inset figures of Fig.~\ref{pss_SALI_LE}a, we present two
typical orbits in the $(x,y)$--plane of the standard model. One is regular
(bottom left inset), with initial condition:
$(x,y,p_{x},p_{y})=(0,-0.625,-0.201,-0.06)\quad (\text{orbit R})$ and the other
chaotic (bottom right inset), with initial condition:
$(x,y,p_{x},p_{y})=(0,-0.625,-0.002,-0.24)\quad (\text{orbit C}),$ both for the
Hamiltonian value $H=-0.360$. Their qualitatively different behavior is also
shown in the PSS in Fig.~\ref{pss_SALI_LE}a ($(y,p_{y})$--plane). The chaotic
orbit C (gray points) tends to fill with scattered points the chaotic region,
while the ordered orbit R (black points) creates a set of points that form a
closed invariant curve, on the left part of the picture.
In the top inset figure of Fig.~\ref{pss_SALI_LE}a, we show the different
morphologies of three regular orbits around the periodic orbits in the centers
of the three largest islands of regular motion present in the associated PSS.
The model's basic barred shape, in the $(x,y)$--plane is mainly provided by
trajectories around the main family of periodic orbits, the so--called x1
family (following the nomenclature of \citealt{ConPa}) which are orbits
elongated along the bar, i.e along the $x$--axis in our case, and inside
corotation. The stability island on the right gives rise to elliptical--like
orbits, around the x2 family, elongated perpendicular to the bar, while the
island on the left contains orbits which are elliptical-like but retrograde,
very slightly elongated perpendicular to the bar and around the family x4.

We then apply the SALI method to the orbits R and C and present their different
typical evolution behaviors: For the chaotic orbit C (gray curve in
Fig.~\ref{pss_SALI_LE}b), the SALI tends to zero $(\simeq10^{-16})$
exponentially after some transient time, while for the regular orbit R, it
fluctuates around a positive number (black curve in Fig.~\ref{pss_SALI_LE}b).
The corresponding mLCE and $\sigma_1$, for these orbits is shown in
Fig.~\ref{pss_SALI_LE}c, where the $\sigma_1$ for the regular orbit tends to
zero, while for the chaotic orbit it tends to a positive number after a long
integration time.

This comparison shows clearly an advantage of SALI (GALI$_2$), namely its
ability to detect the chaotic character of an orbit faster than by simply
following the mLCE, which typically takes a long time to converge. Even in our
two DOF model, SALI starts to decrease exponentially after a relatively short
time (in Fig.~\ref{pss_SALI_LE}b, $t\leq10^{3}$) while the corresponding mLCE
$\sigma_1$ starts to converge to some positive non-zero value for $t\geq10^{3}$
(see gray curve in Fig.~\ref{pss_SALI_LE}c).

Exploiting  the efficiency of SALI, we take initial conditions on the
$(y,p_{y})$--plane (with $x=0$) and calculate the values of the index to detect
very small regions of stability (or instability) more globally. We are thus
able to construct a map of chaotic and regular regions, very similar to what is
depicted in a PSS, but with more accuracy and higher resolution (albeit at a
somewhat longer CPU time). Furthermore, as a byproduct of our application, we
obtain an accurate estimate of the percentage of regular to chaotic orbits on a
surface of section of the given energy.

\begin{figure*}
\epsfig{file=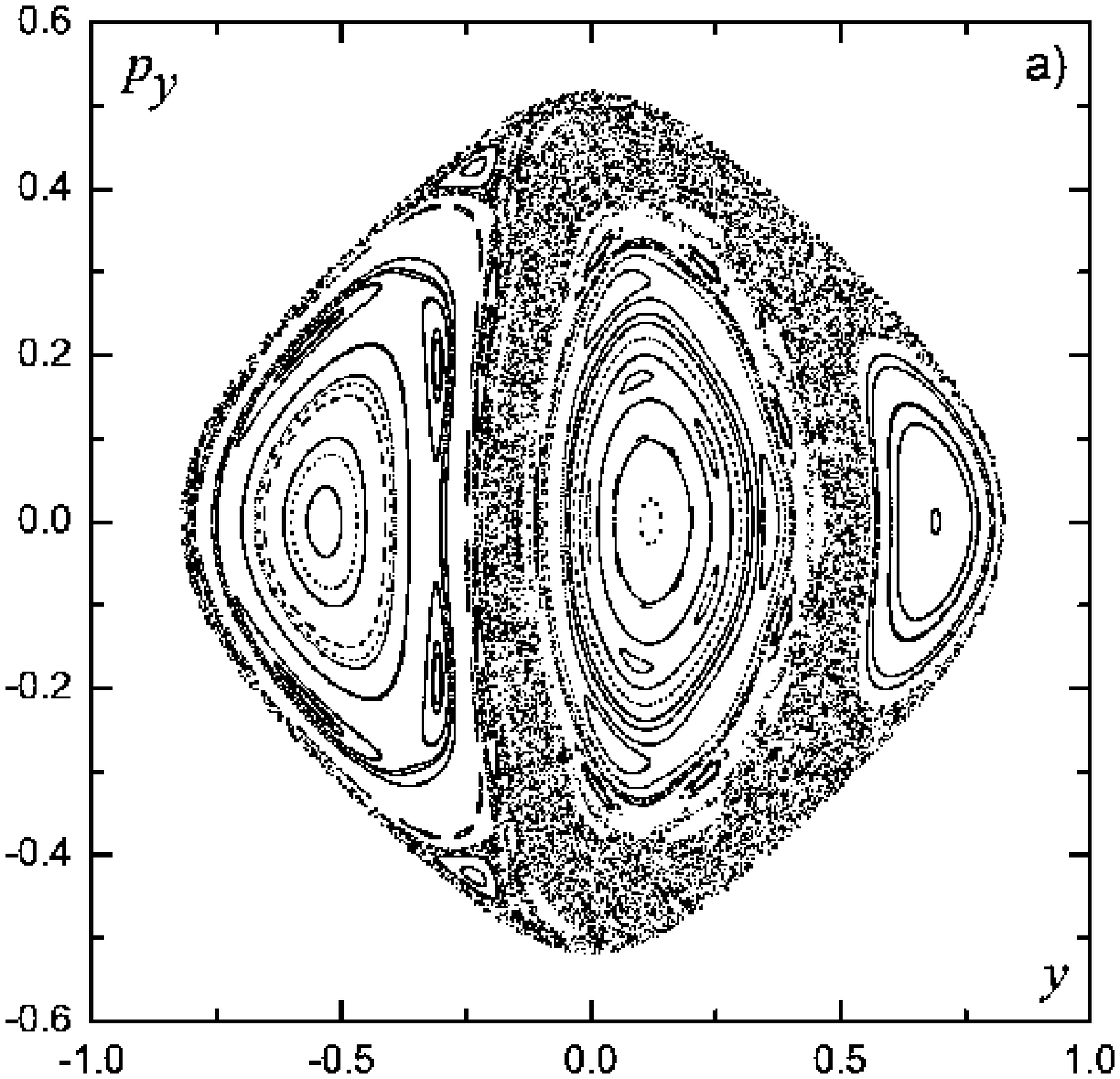,height=5.cm,width=7.5cm}\hspace{0.75cm}
\epsfig{file=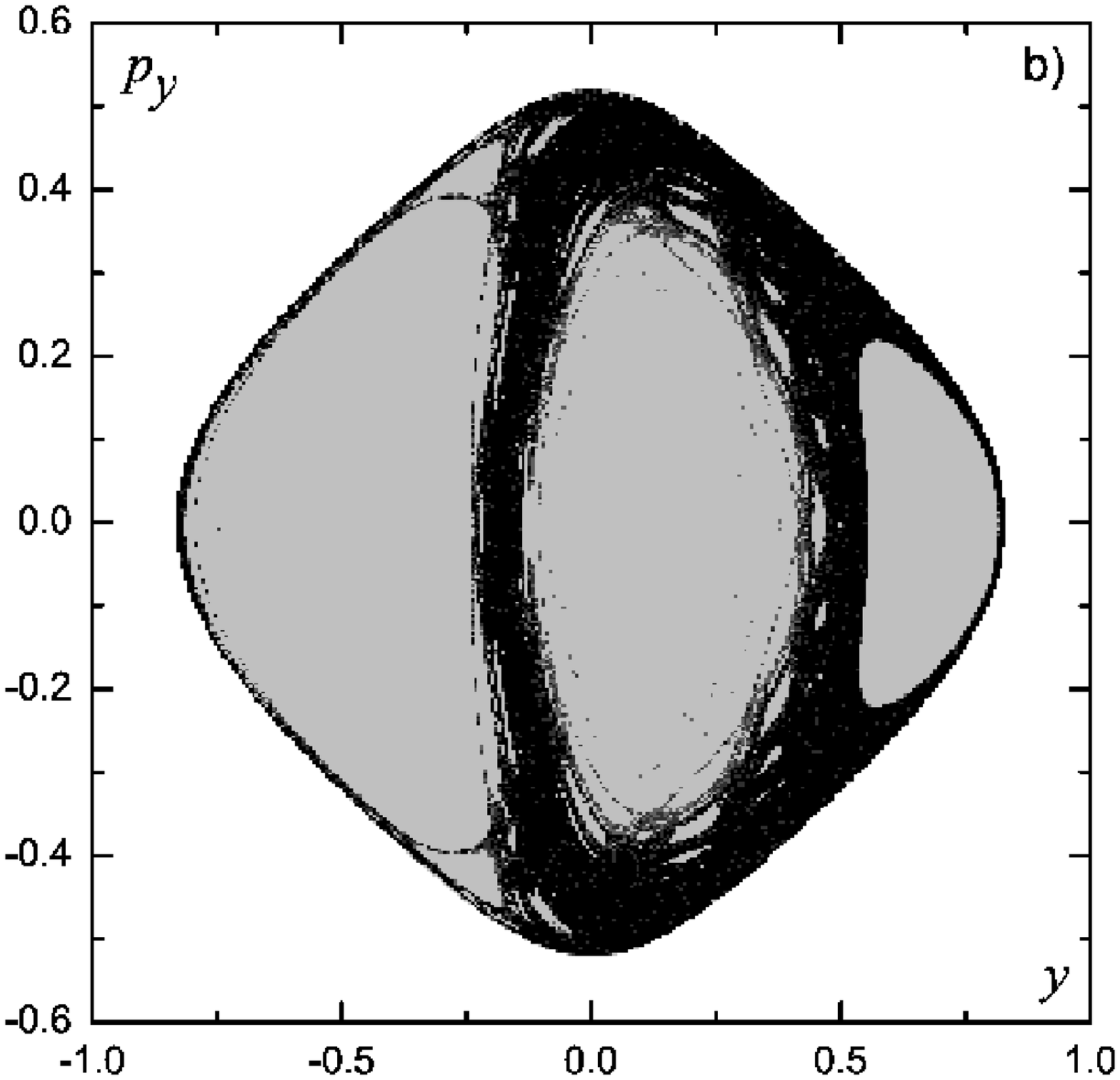,height=5.cm,width=7.5cm}\\\vspace{0.25cm}
\epsfig{file=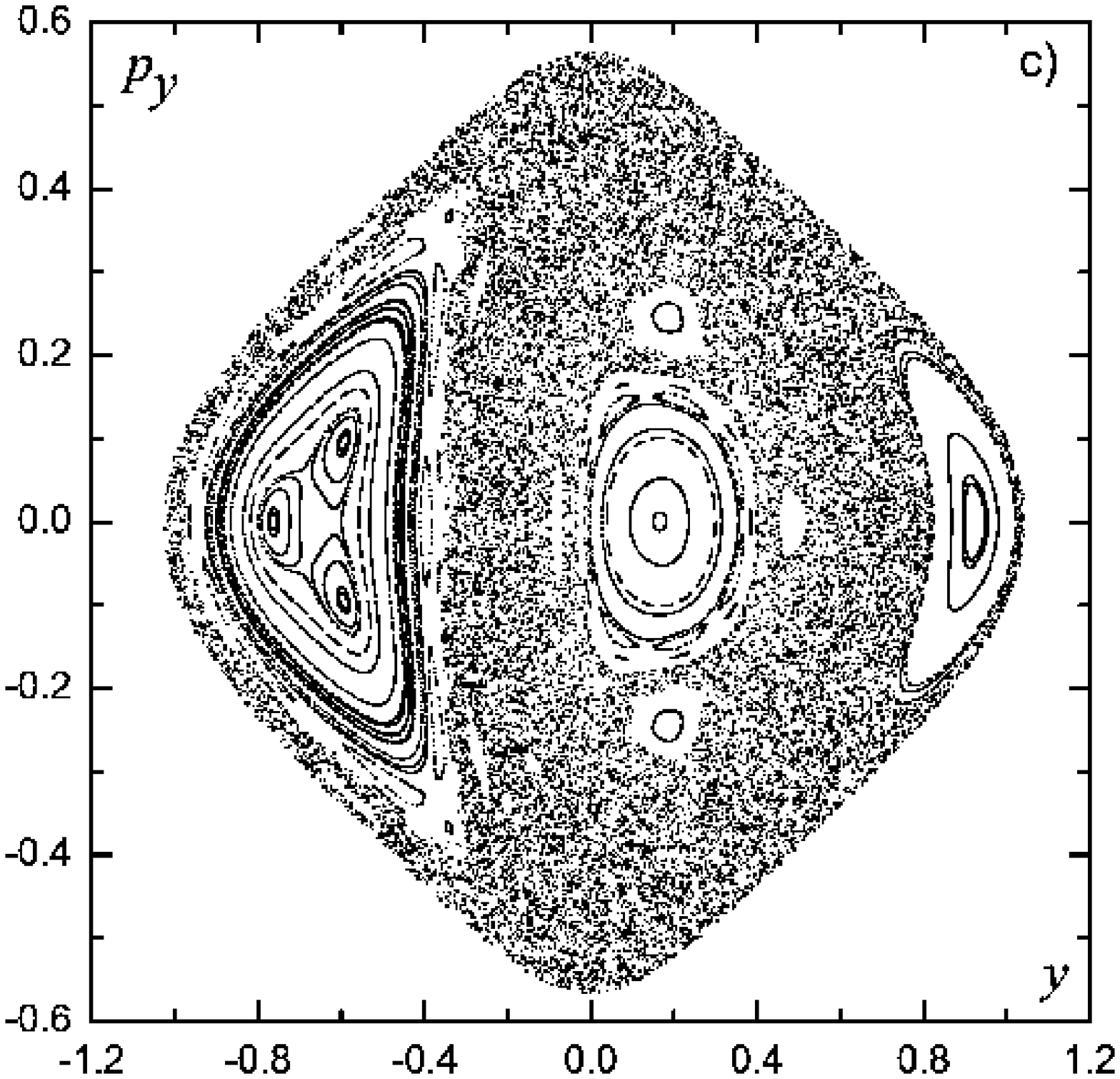,height=5.cm,width=7.5cm}\hspace{0.75cm}
\epsfig{file=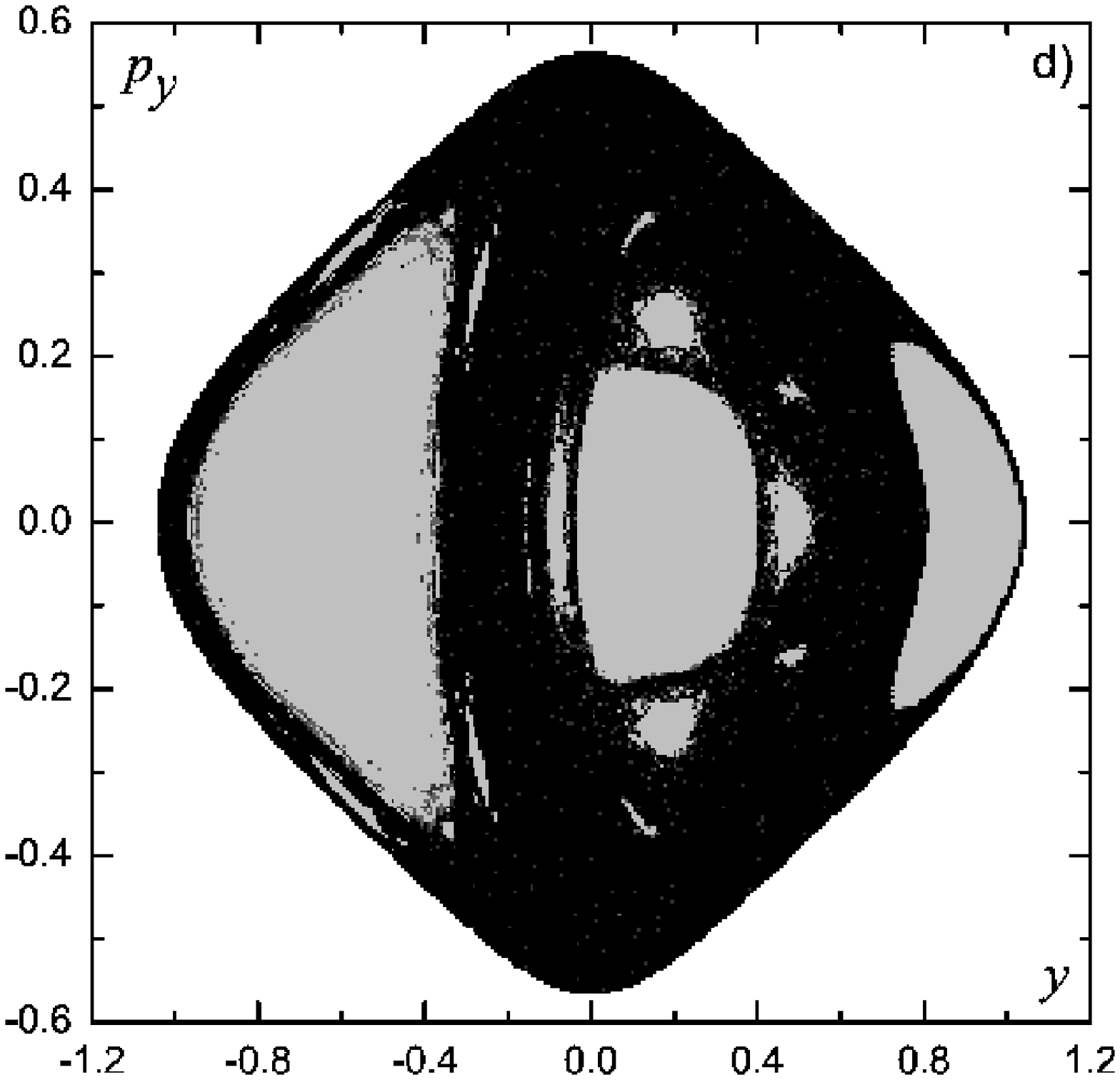,height=5.cm,width=7.5cm}\\\vspace{0.25cm}
\epsfig{file=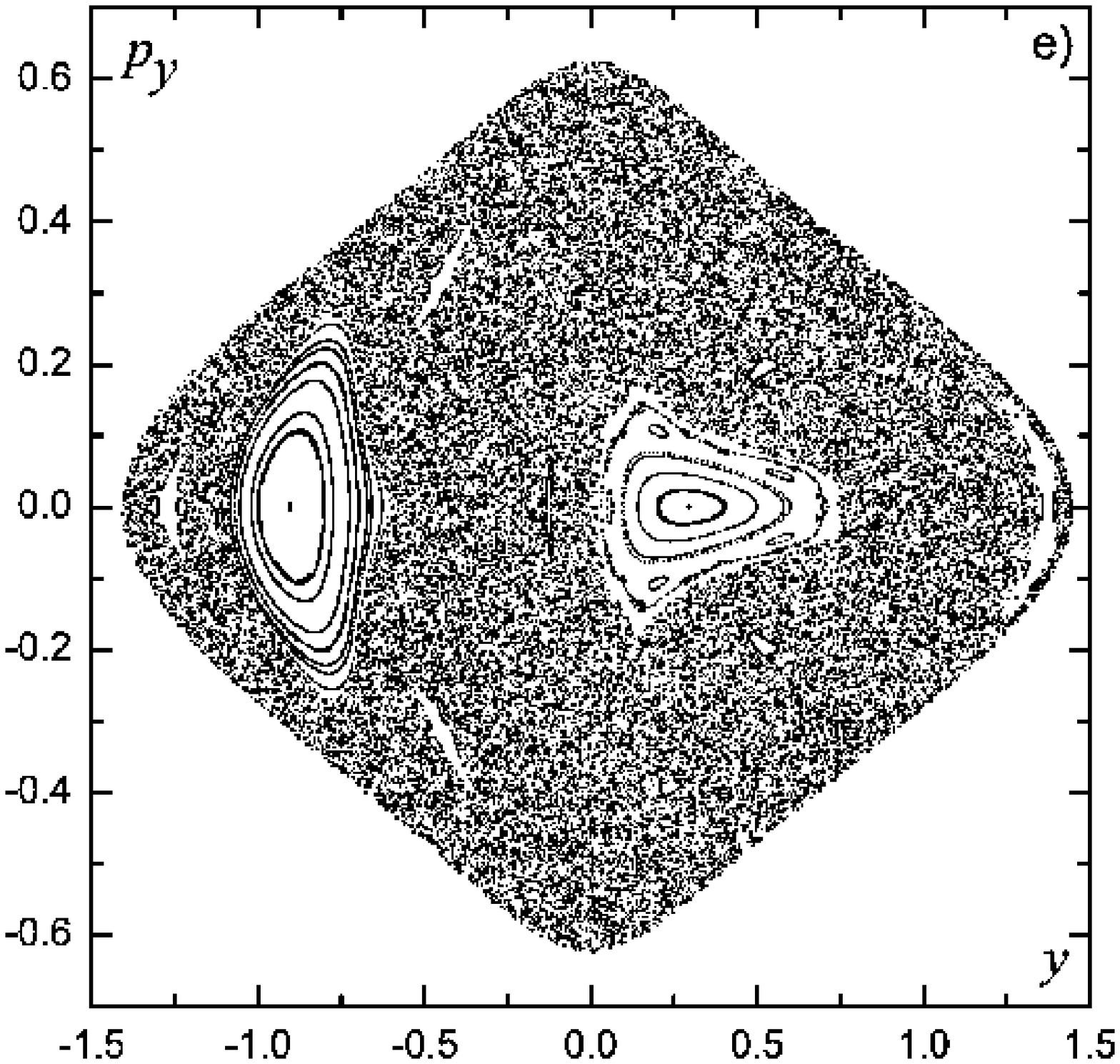,height=5.cm,width=7.5cm}\hspace{0.75cm}
\epsfig{file=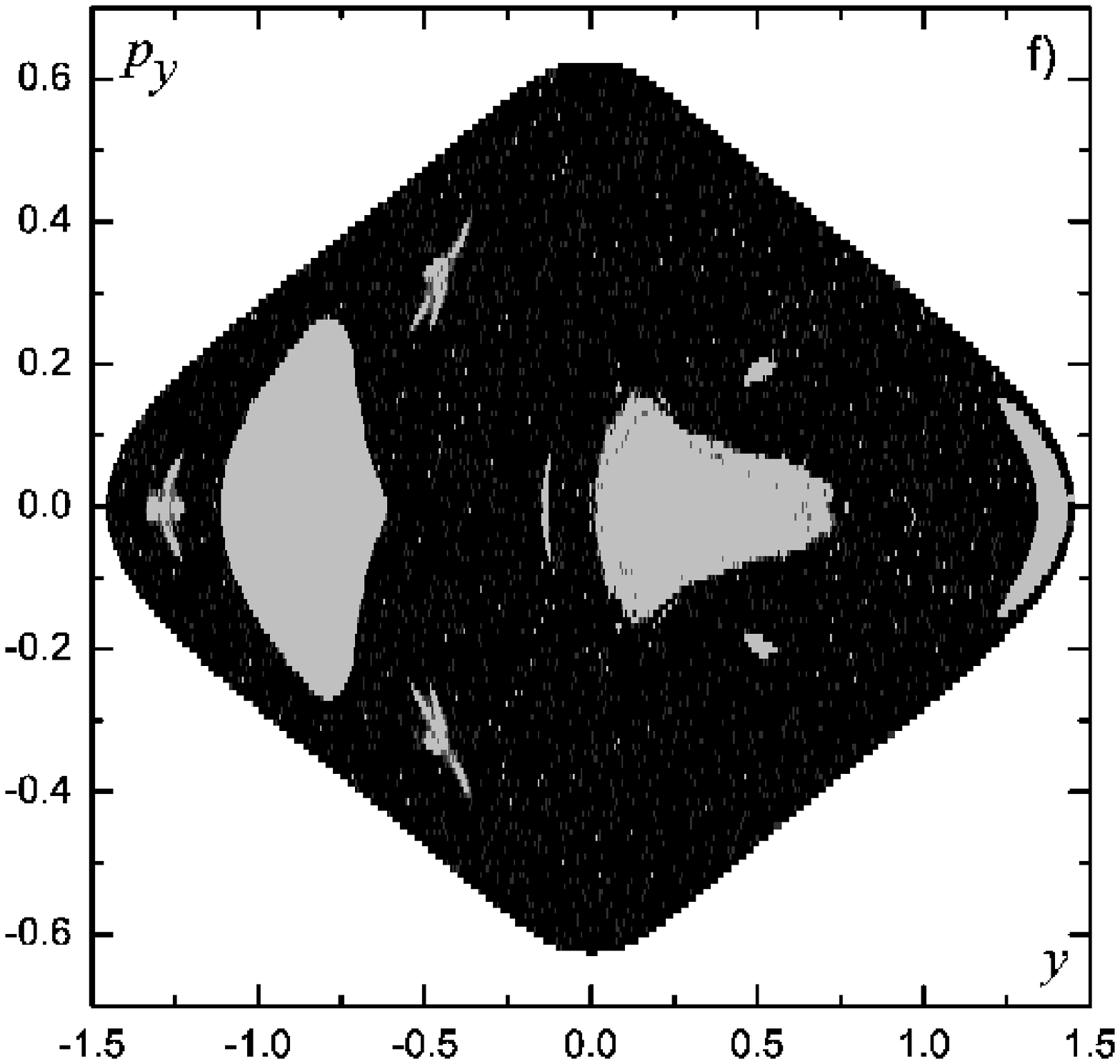,height=5.cm,width=7.5cm}\\\vspace{0.25cm}
\epsfig{file=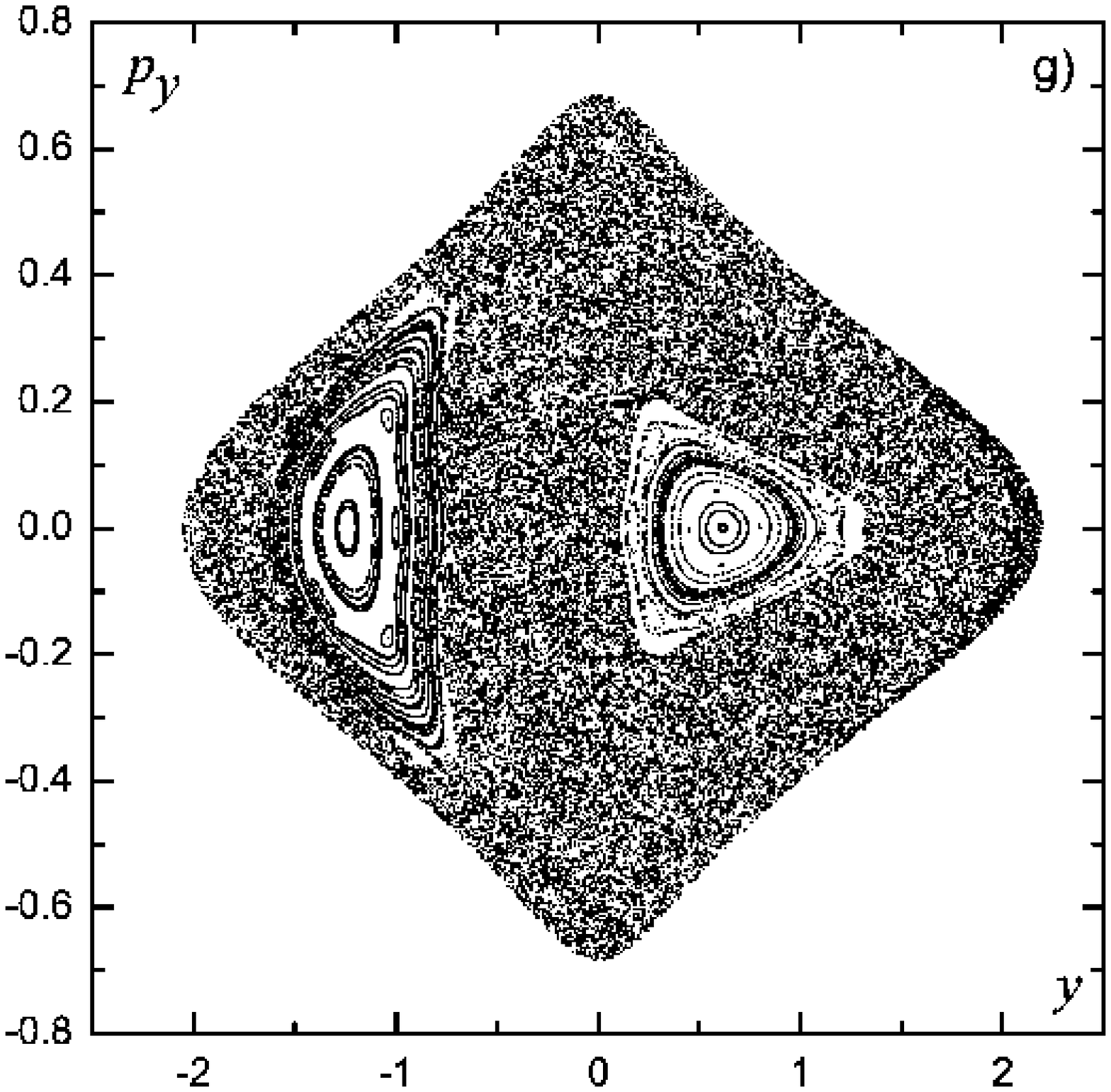,height=5.cm,width=7.5cm}\hspace{0.75cm}
\epsfig{file=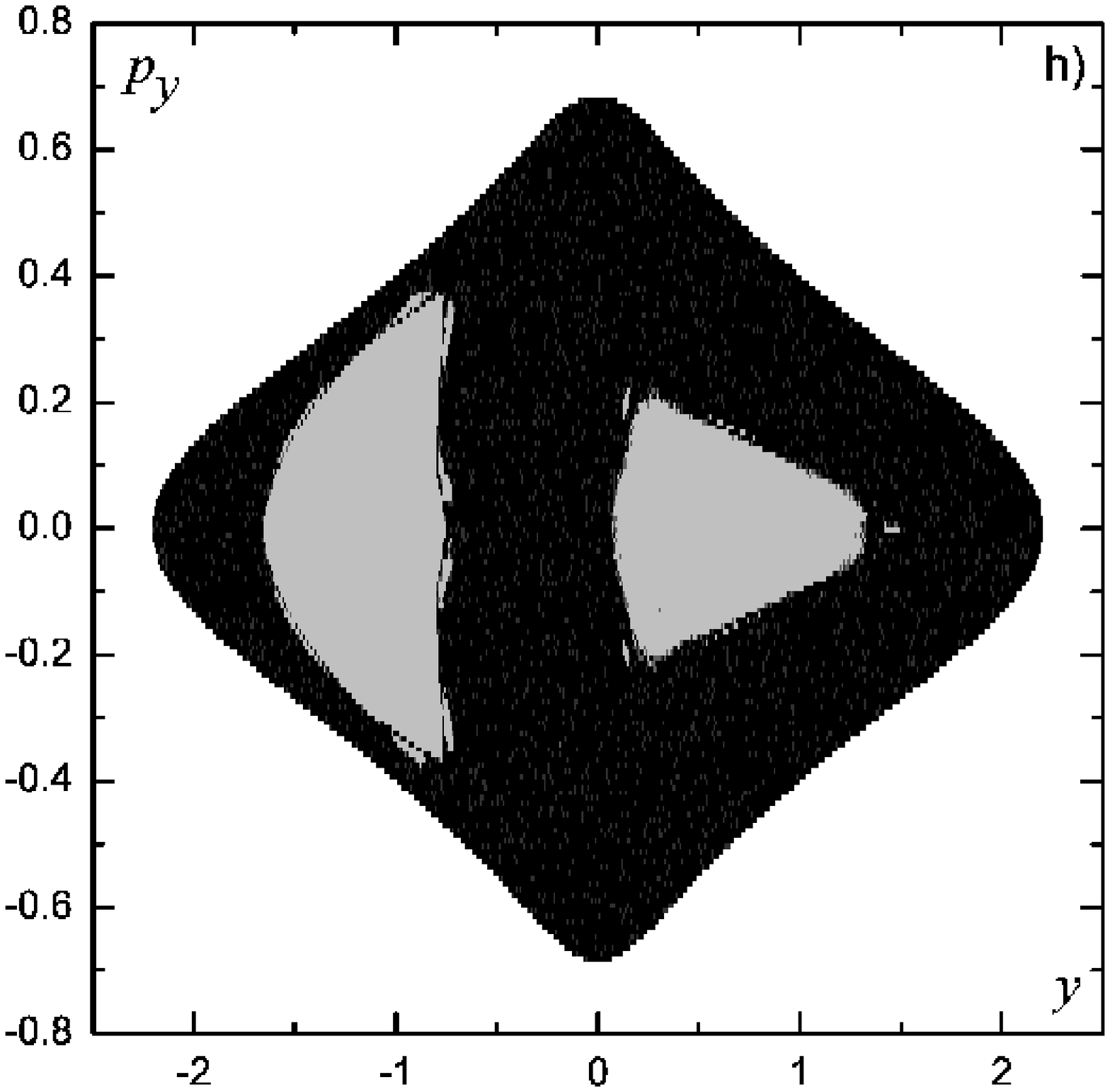,height=5.cm,width=7.5cm}\\
 \caption{The PSS (left column) and SALI (right column) methods are in good agreement
 when surveying the same projection of the phase space. In the left column we show the
 PSS for the  two DOF model potential with $H=-0.360$ (panel a), $H=-0.335$ (panel c),
 $H=-0.300$ (panel e) and $H=-0.260$ (panel g). In the right column we present regions of
 different values of the SALI for 50,000 initial conditions on the
 $(y,p_{y})$--plane for the same values of the Hamiltonian (panels b,d,f and h,
 respectively). The light gray colored areas correspond to regular orbits, while
 the dark black ones to chaotic. Note the excellent agreement between the two
 methods as far as the gross features are concerned, as well as the fact that
 the SALI can easily pick out small regions of stability which the PSS has
 difficulties detecting, like those around the central islands in panels b,d,f
 and h.} \label{SALI_pss_ener_a}
\end{figure*}

\begin{figure}\centering
  \includegraphics[height=0.275\textheight]{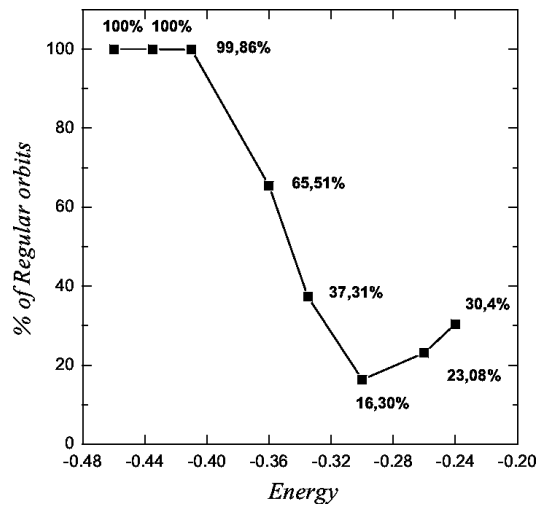}
  \caption{Percentages of regular orbits for several values of the
  energy ($H=$ -0.460,-0.435,-0.410,-0.360,-0.335,-0.300,-0.260,-0.240)
  in the two DOF Ferrers model. Note that while the fraction of the
  regular orbits decreases
  sharply as the energy grows above $H=$ -0.4, this tendency is reversed at higher
  energy values.}
  \label{per_EN_2dof}
\end{figure}

For example, choosing different values of the energy and using a sample of
50,000 initial conditions equally spaced on the same $(y,p_{y})$--plane, we
plot on the left column of Fig.~\ref{SALI_pss_ener_a} the PSS for $H= -0.360,
-0.335, -0.300, -0.260$ and on the right the corresponding final SALI values
obtained from the selected grid of initial conditions, where each point is
colored according to its SALI values at the end of the integration. In the SALI
plots, the light gray color corresponds to regular orbits, the black color
represents the chaotic orbits/regions, while the intermediate gray shades
between the two represent orbits with small rate of local exponential
divergence, and/or orbits whose chaotic nature is revealed after long times,
like for example the so-called sticky orbits, i.e. orbits that "stick" onto
quasiperiodic tori for long times. Note that the fraction of these orbits
(which lie mainly around the borders of the islands of stability) is very small
(few $\%$ of the total amount of initial conditions) and can be discerned by
eye in Fig.~\ref{SALI_pss_ener_a} only if one focuses on these particular
regions.
Thus, calculating from all these initial conditions the percentage of regular
orbits, we are able to follow how this fraction varies as a function of the
total energy. This problem was already addressed for four models by
\cite{ABMP} but with fewer orbits, i.e. larger error bars. In the present
model, the chosen energy values start from a regime of total order, at
$H=-0.46$, up to values near the escape energy, $H_{esc}=-0.20$. The
distinction between chaotic and regular orbit is that SALI$<10^{-8}$ for
chaotic orbits and $\geq10^{-8}$ for regular ones. In the latter range one
does, of course, include ``sticky" chaotic orbits as well. As is clear from
Fig.~\ref{per_EN_2dof}, although the percentages of the regular orbits
decreases sharply as the energy grows above $H=-0.4$, this tendency is reversed
at higher energy values. Clearly, for $H>-0.3$, despite the fact that many
stability islands have vanished, the size of the ordered domain increases
because the main big island on the left of the PSS (see
Fig.~\ref{SALI_pss_ener_a}f,h) which corresponds to retrograde orbits, has
become larger.
\section{The three DOF model potential}
\label{3dof}

We now turn to the three DOF model and begin a study of ordered and chaotic
domains. Before discussing the more general results about the study of the
phase space of the model, we first present in detail the behavior of our chaos
indicators for some typical orbits. Let us choose a regular and a chaotic orbit
and compare the efficiency of the GALI method with that of the mLCE's.

\begin{figure}\centering
\includegraphics[height=0.245\textheight]{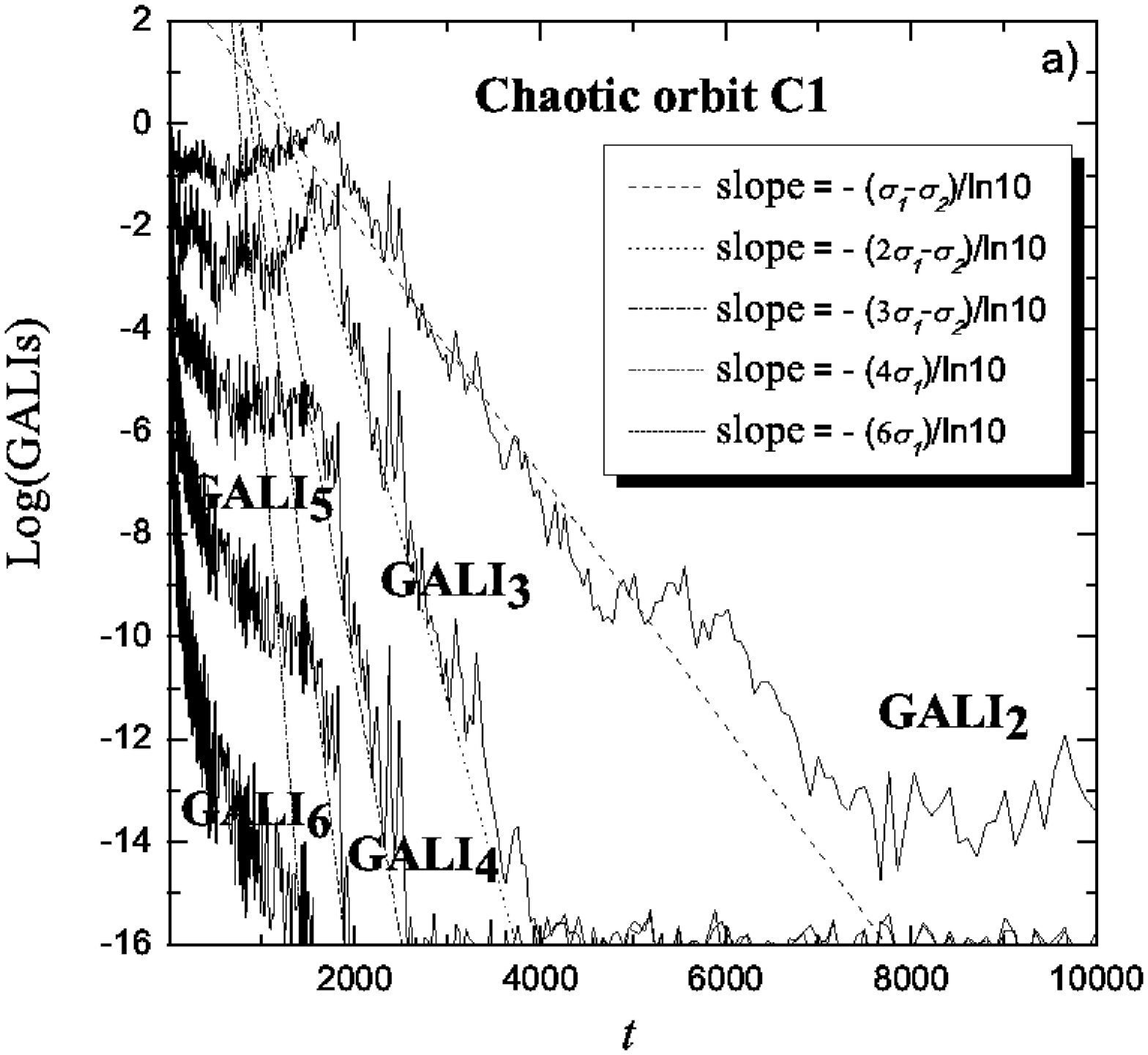}
\includegraphics[height=0.245\textheight]{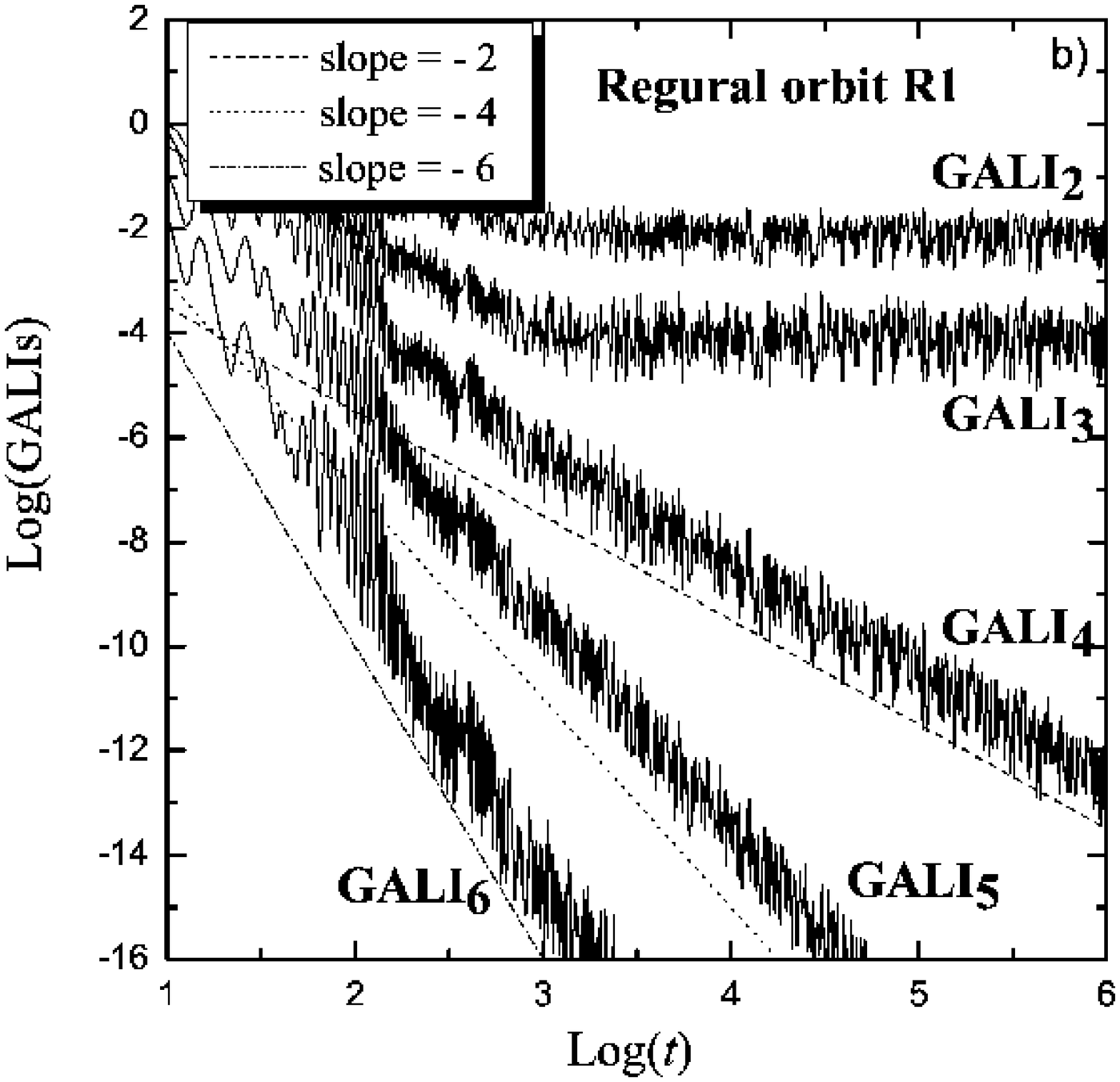}
\includegraphics[height=0.245\textheight]{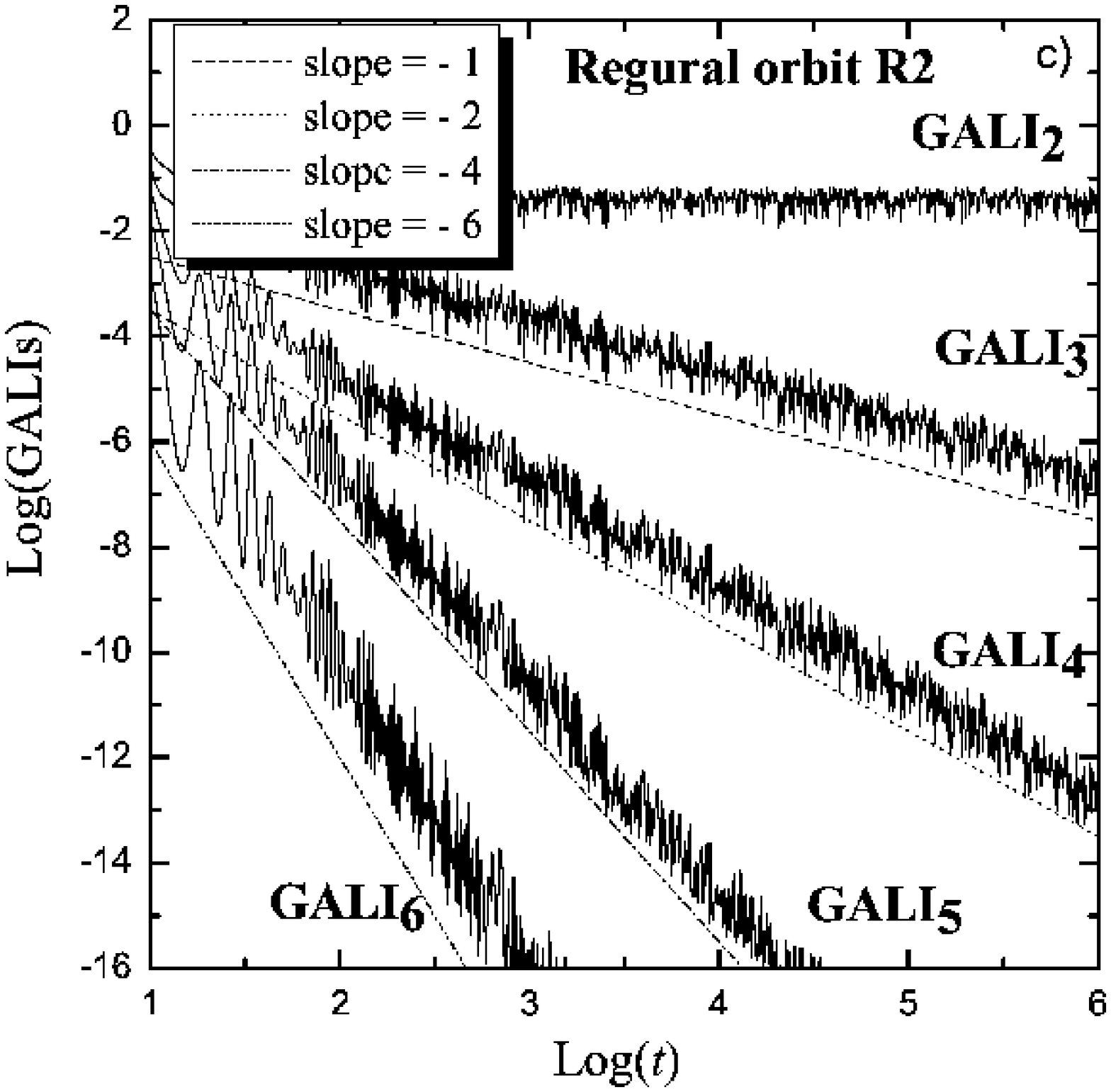}
\caption{Distinguishing chaotic from regular motion with GALI method. a)
Exponential decay of all GALIs for the chaotic orbit C1 (lin-log scale) with
the predicted slopes related to its two largest Lyapunov exponents ($\sigma_1
\approx 0.00910$ and $\sigma_2 \approx 0.00345$). b) Slopes of GALIs for the
regular orbit R1 showing that its motion lies on a 3D torus, where GALI$_{2,3}
\propto$ constant while the GALI$_{4,5,6}$ decay following a power law. c)
Slopes of GALIs for the regular orbit R2 lying on a 2D torus with only
GALI$_2\propto$ constant. } \label{SALI_3dof}
\end{figure}

In Fig.~\ref{SALI_3dof}a,b we show the behavior of a chaotic orbit (C1), with
initial condition: $(x,y,z,p_x,p_y,p_z)=(0.5875,0.0,0.33333,0.0,0.2,0.0)$  and
of a regular orbit (R1), with initial condition:
$(x,y,z,p_x,p_y,p_z)=(0.97917,0,0.04167,0,-0.17778,0)$. For the chaotic one,
the GALI$_2$ (or SALI) has become almost zero for $t \simeq 10^4$. However, as
it is clear in Fig.~\ref{SALI_3dof}a, the higher order GALI$_k, \ k=5,6$,
indicate the chaoticity of the orbit already by $t \simeq 10^3$, since they
have reached very small values at that time. Note the good agreement between
the predicted slopes related to its two largest Lyapunov exponents ($\sigma_1
\approx 0.00910$ and $\sigma_2 \approx 0.00345$) given by Eq.~(\ref{GALI:1}).
In Fig.~\ref{SALI_3dof}b we show the GALI indices for the regular orbit R1,
where both GALI$_{2,3} \propto$ constant and the GALI$_{4,5,6}$ decay following
the power laws:
\begin{flalign}\label{GALI_reg_3d}
\text{GALI}_4(t)\propto \frac{1}{t^2}, \ \text{GALI}_5(t) &\propto \frac{1}{t^4},\ \text{GALI}_6(t)\propto \frac{1}{t^6},
\end{flalign}
obtained from Eq.~(\ref{GALI:2}) for $m=3$, indicating regular motion which lies on a 3D torus.

One example of a low dimensional motion (lower than 3) is provided by the
regular orbit (R2), with initial condition:
$(x,y,z,p_x,p_y,p_z)=(0.5875,0.0,0.29770,0.0,0.33750,0.0)$. In
Fig.~\ref{SALI_3dof}c, we show the behavior of the Log(GALI) indices and their
slopes for the regular orbit R2. Note that only GALI$_2$ remains constant while
the GALI$_{3,4,5,6}$ tend to zero following power laws predicted by
Eq.~(\ref{GALI:2}):
\begin{flalign}\label{GALI_reg}
 \text{GALI}_2(t) &\propto const., \quad  \text{GALI}_3(t)\propto \frac{1}{t},\quad \text{GALI}_4(t)\propto
 \frac{1}{t^2}, \nonumber \\
 \text{GALI}_5(t) &\propto \frac{1}{t^4},\qquad  \ \text{GALI}_6(t)\propto \frac{1}{t^6},
\end{flalign}
for $m=2$. This implies that the orbit's motion lies on a 2D torus even though
in three DOF Hamiltonian systems one generally expects the dimension of the
torus to be three.

\begin{figure*}\hspace{-0.3cm}
\hspace{0.1cm}
{\includegraphics[height=0.225\textheight]{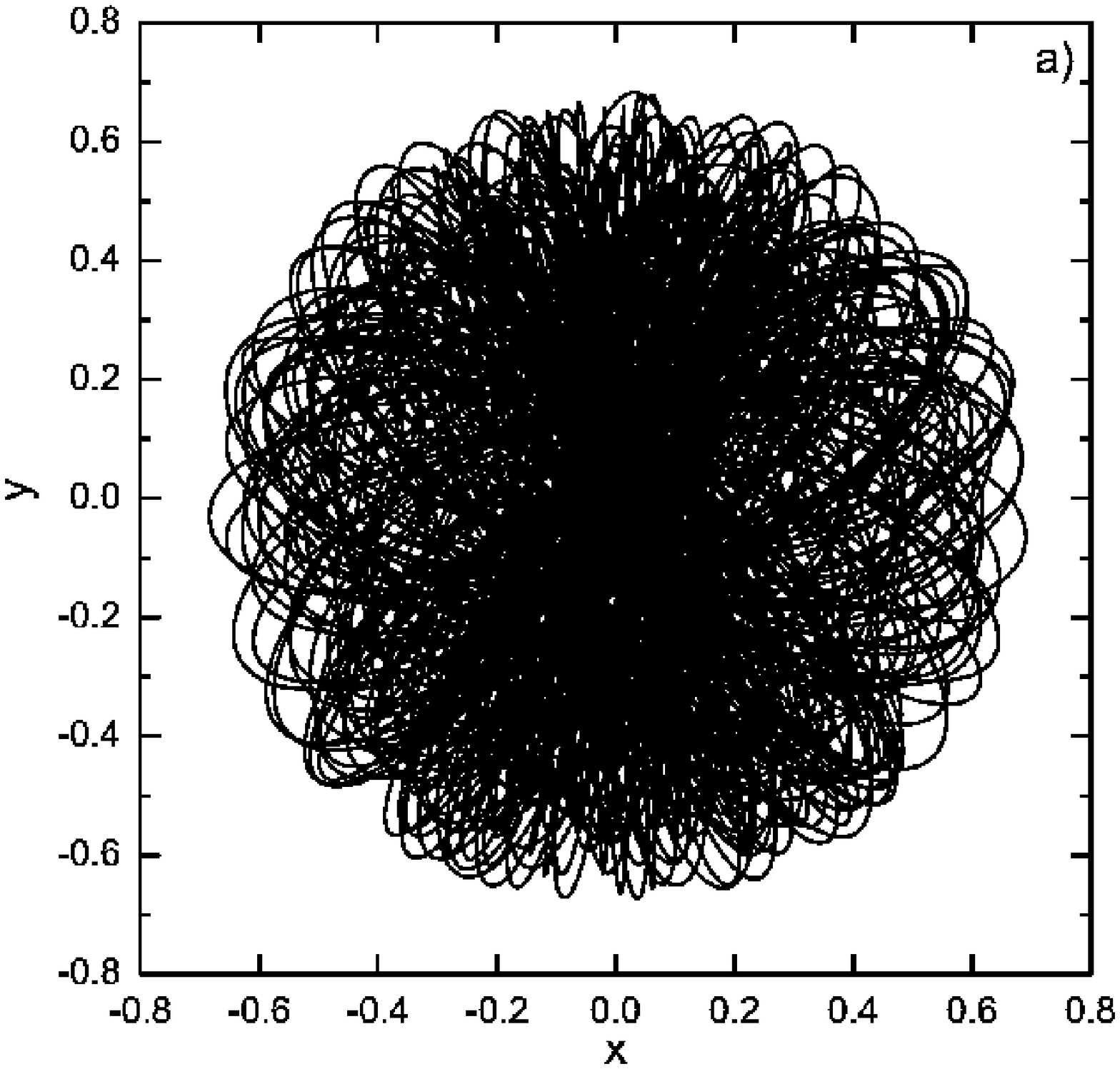}}\hspace{0.1cm}
{\includegraphics[height=0.225\textheight]{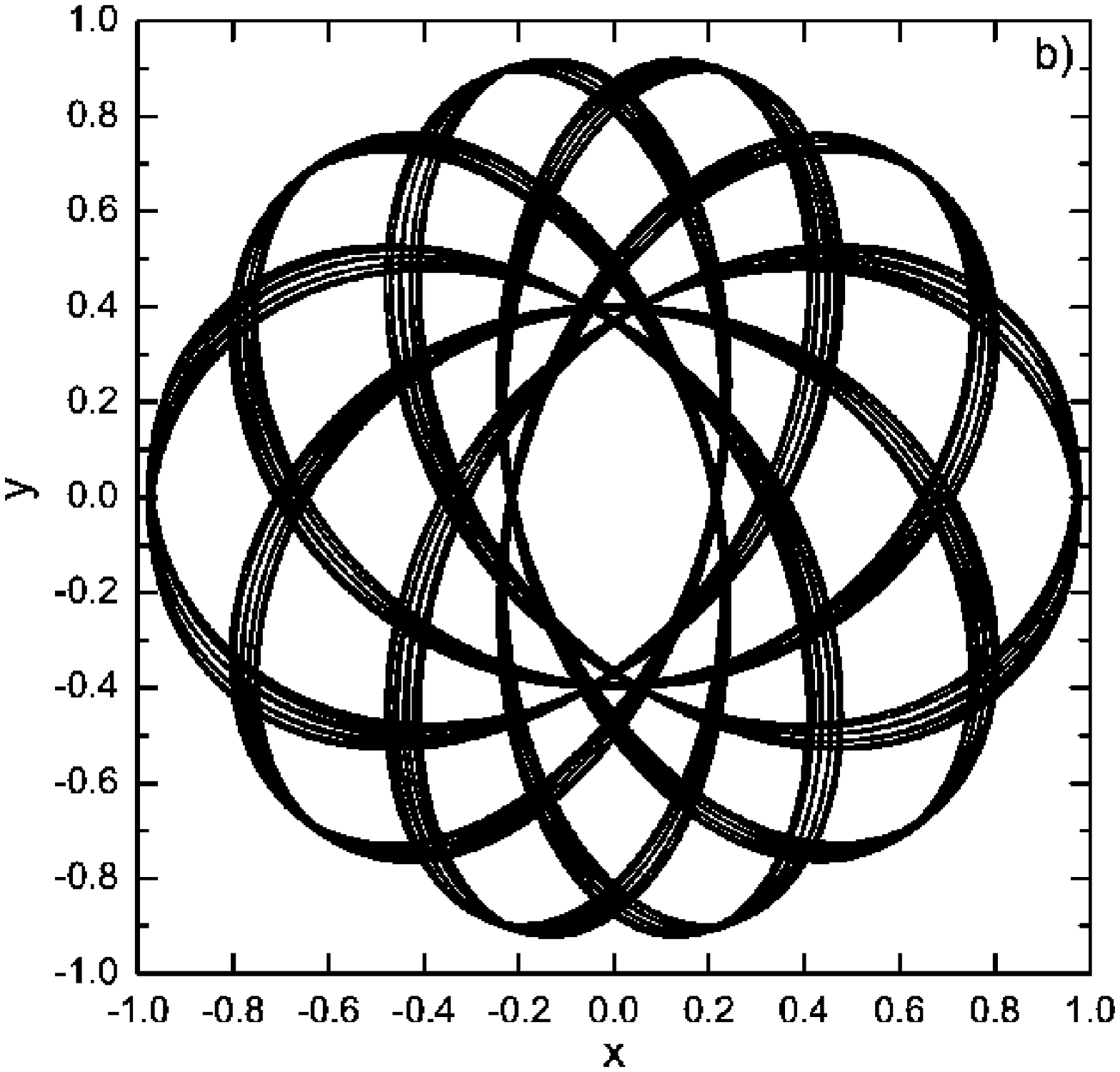}}\hspace{0.1cm}
{\includegraphics[height=0.225\textheight]{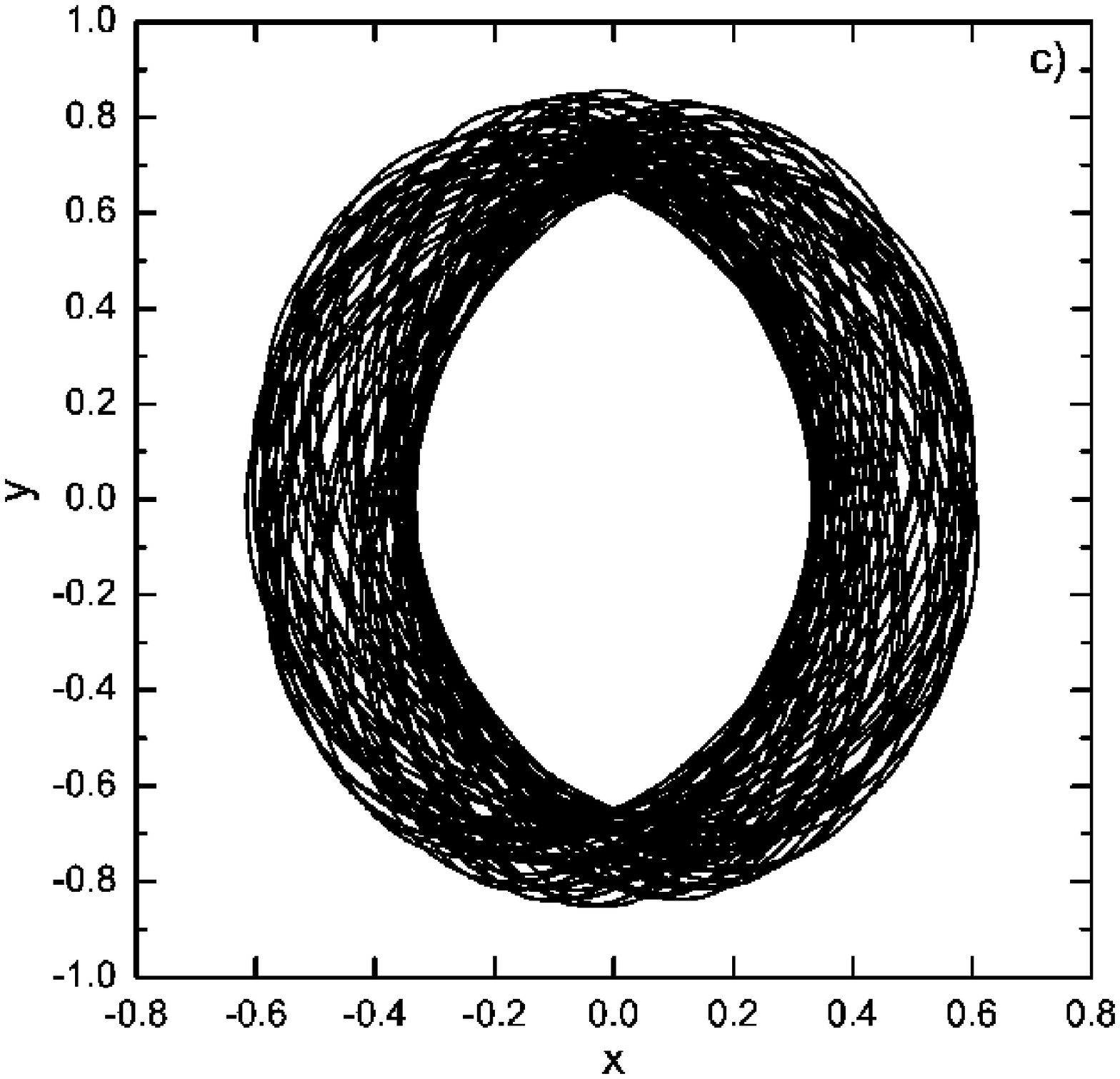}}\\
{\includegraphics[height=0.225\textheight]{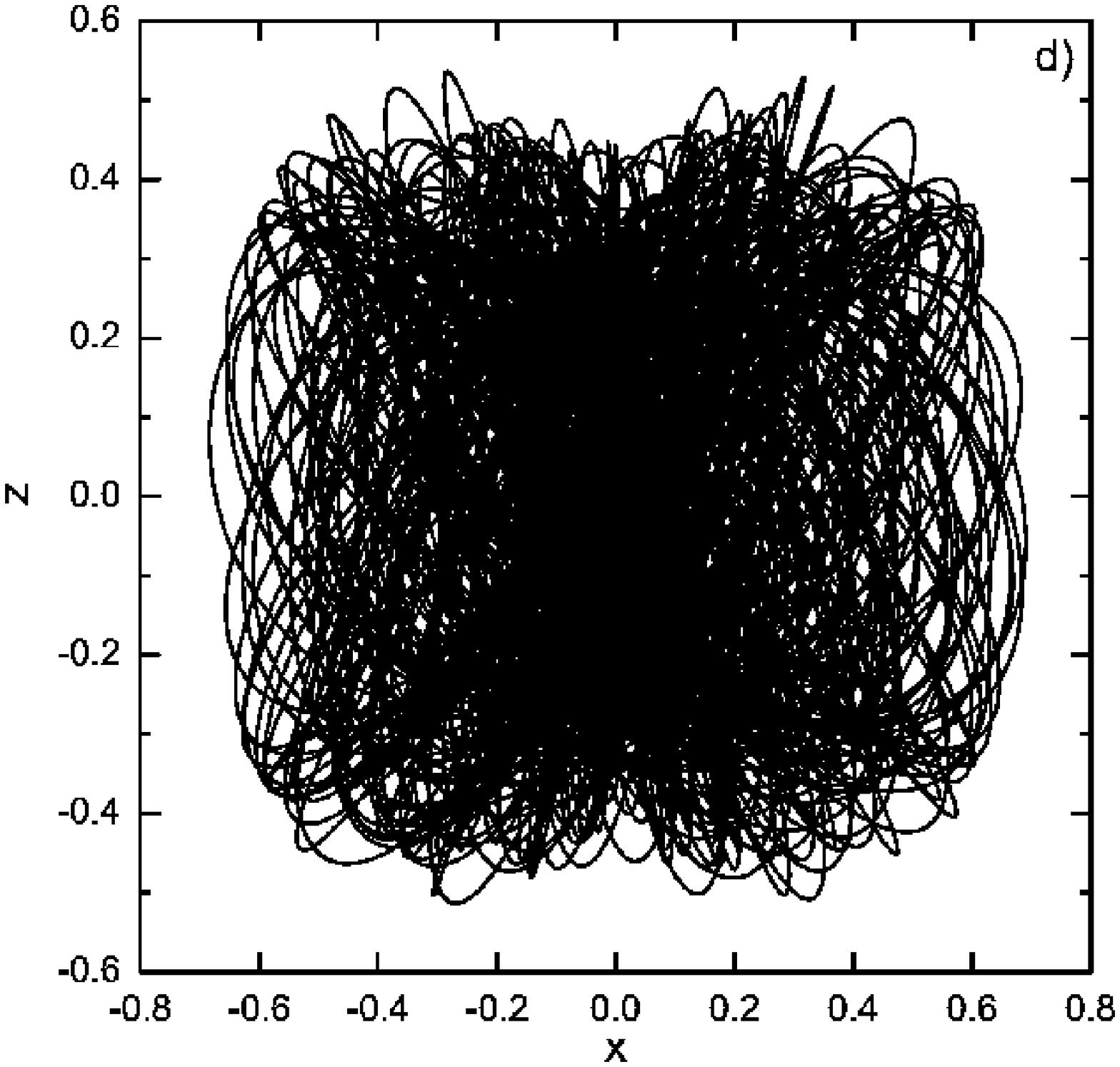}}
{\includegraphics[height=0.225\textheight]{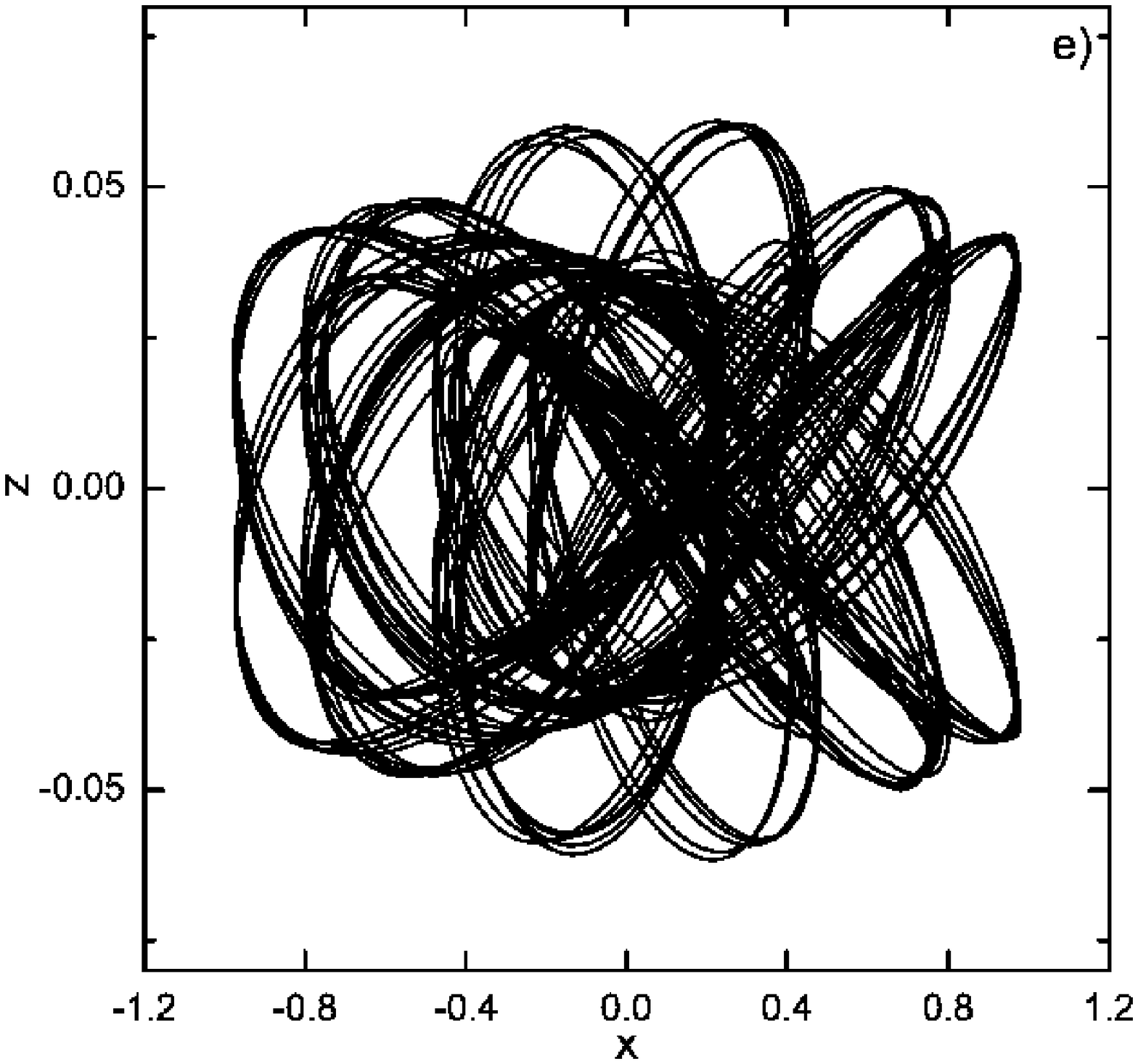}}
{\includegraphics[height=0.225\textheight]{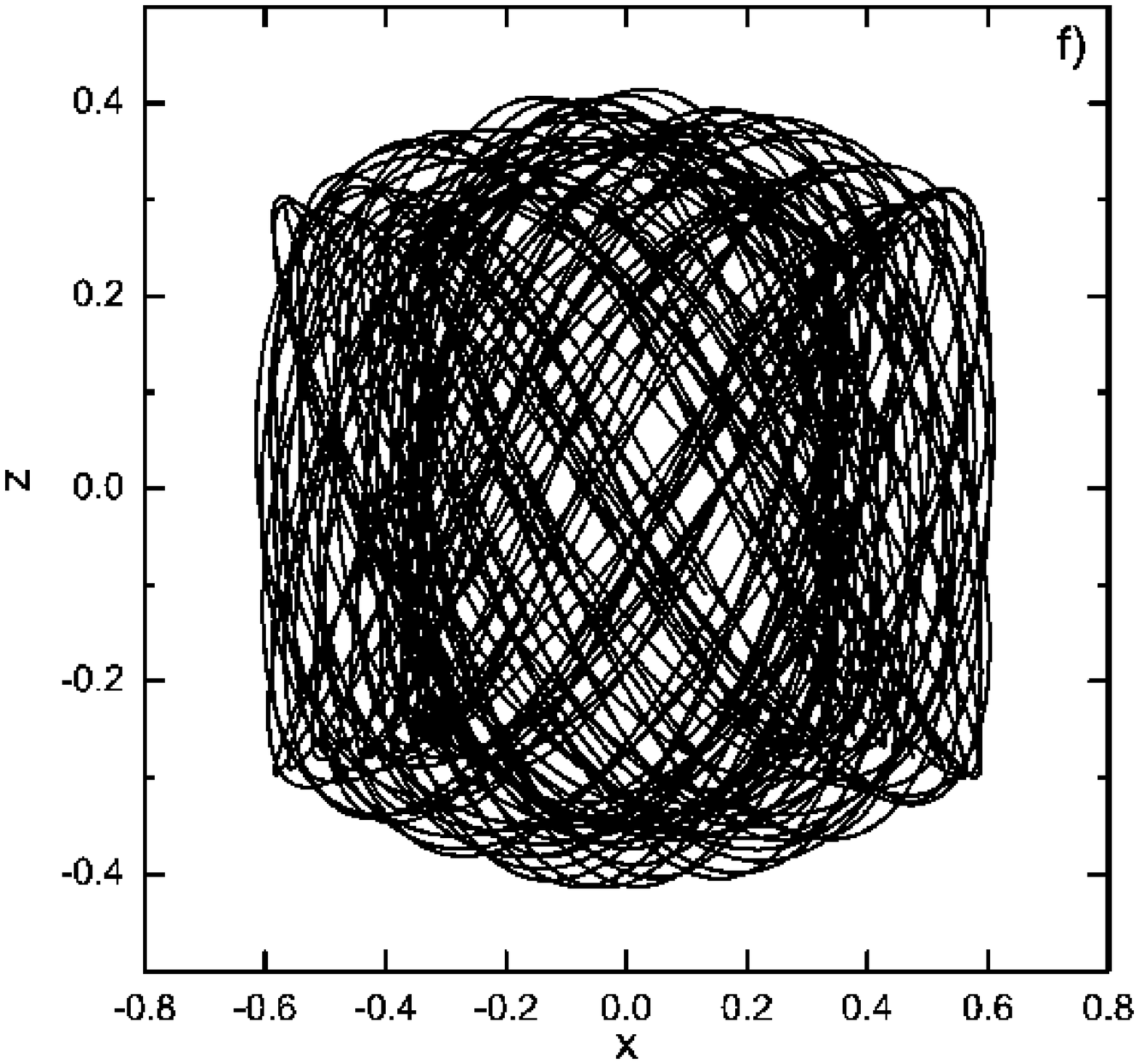}} \caption{Two projections
($(x,y)$ and $(x,z)$--planes) of the chaotic orbit C1 (1st column) and the
quasiperiodic orbits R1 (2nd column) and R2 (3rd column). Note the different
``complexity'' between R1, whose regular motion lies on a 3D torus, and R2,
which lies on a 2D as GALI indices detected.}
 \label{2_3D_torus_freq}
\end{figure*}

In Fig.~\ref{2_3D_torus_freq} we show the corresponding $(x,y)$ and $(x,z)$
projections for the above C1 (1st column), R1 and R2 orbits (2nd and 3rd column
respectively). The chaotic one fills up its available regime. Regarding the two
regular ones, there is a clear difference reflected in their projections. The
``complexity" of R1 (quasiperiodic motion on a 3D torus) is more pronounced
than the one of the orbit R2 (quasiperiodic motion on a 2D torus). Thus, in
such cases GALI$_3$ offers an extra advantage in helping us detect different
``degrees" of regularity and does not serve only to distinguish between chaotic
and regular motion.

\section{The distribution of orbits in phase space}
\label{3Dphase_space}

In this section we focus on the detection and quantification of the different
kinds of orbital motion (regular and chaotic) in phase (and configuration)
space, as some parameters of the bar component vary. A similar study, in
different potentials, was done in \cite{ZS1,ZS2}, using the Lyapunov spectra as
a tool. Here we concentrate mainly in the bar component and in understanding
how its different properties can affect the general stability of the system. We
use a relatively large number of trajectories (50,000), more than one set of
initial conditions for the survey of the phase space, and the SALI method as a
chaos detector for the reasons discussed in the previous sections. We discuss
the relative fraction of regular and chaotic motion not only as a function of
its spatial location but also as a function of the total energy and to
correlate it with the bar strength, i.e. the relative non-axisymmetric forcing.
\begin{figure*}
\includegraphics[height=0.25\textheight]{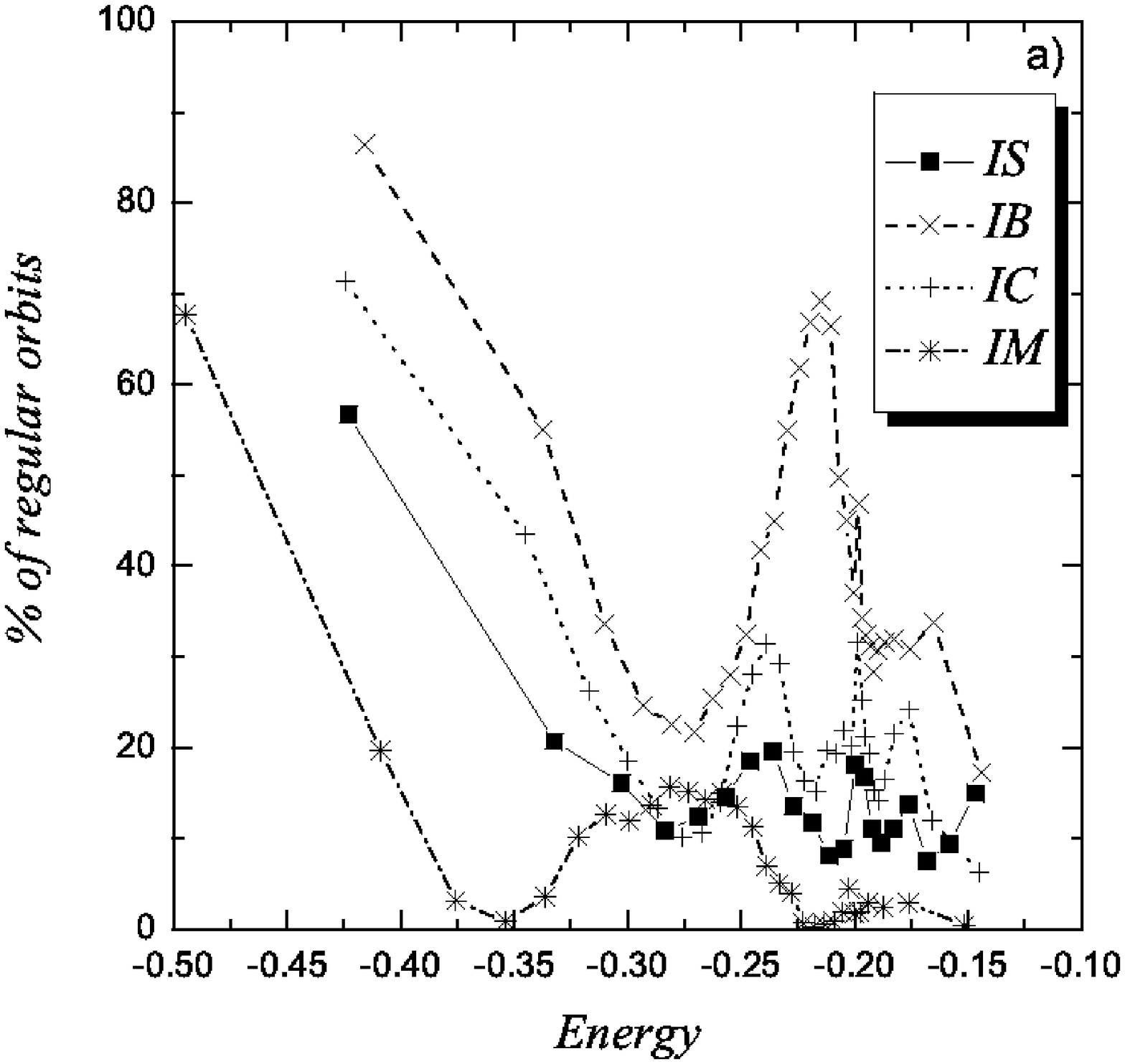}\hspace{1.25cm}
\includegraphics[height=0.25\textheight]{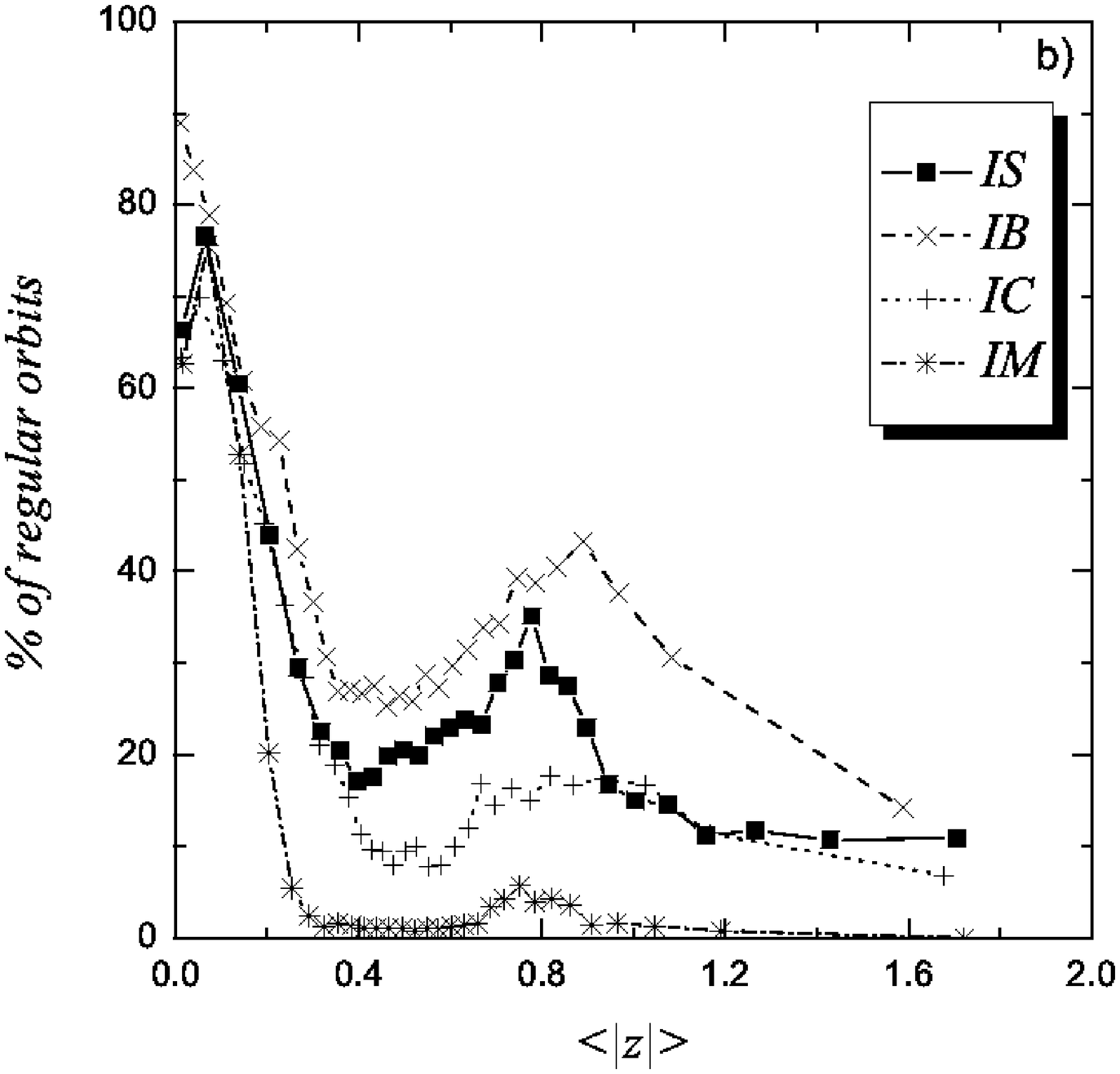}\\
\includegraphics[height=0.25\textheight]{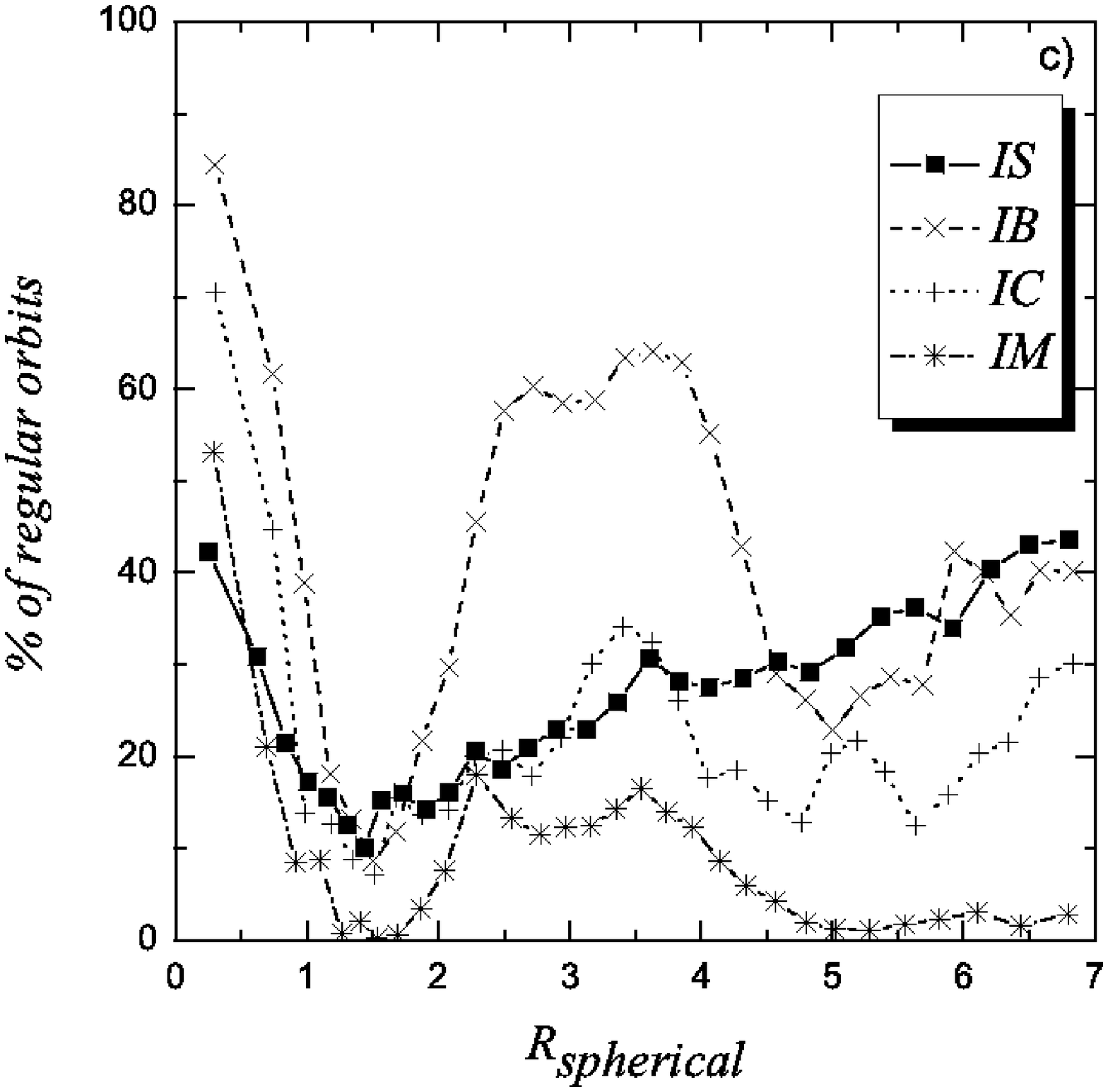}\hspace{1.5cm}
\includegraphics[height=0.25\textheight]{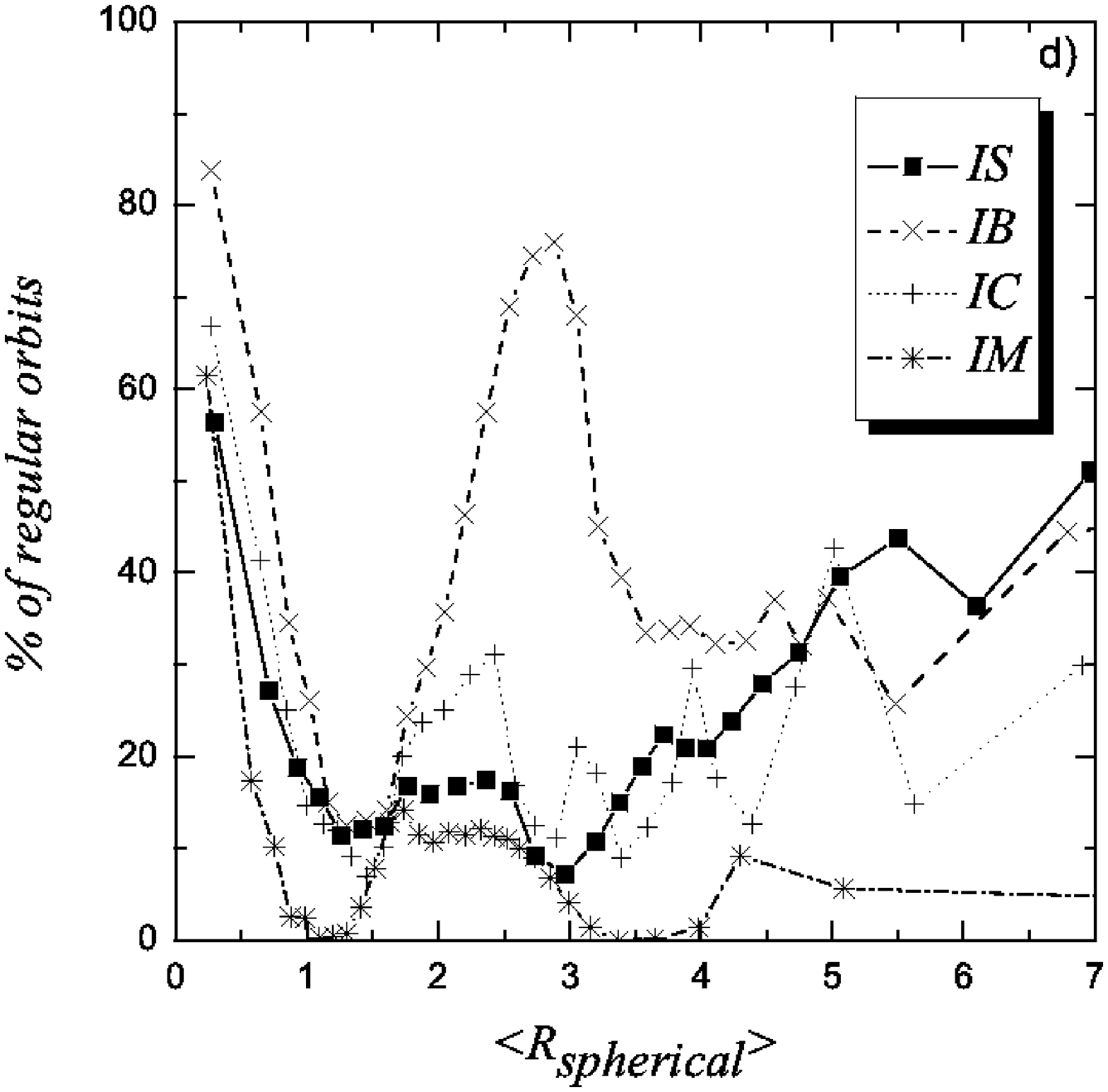}
\caption{Percentages of regular orbits as a function of (a) the energy, (b) the
averaged $<|z|>$ value, (c) the initial spherical radius $R_{spherical}$, and
(d) the averaged $<R_{spherical}>$ radius of the orbit's evolution, for models
$IS$, $IB$, $IC$ and $IM$.} \label{perc_sev}
\end{figure*}
\subsection{Initial conditions}\label{inconcases}

We now need to define the sample of orbits whose properties (chaos or
regularity) we will examine. The best of course would have been to draw these
from a distribution function of the model. This, however, is not available, so
we choose initial conditions on different grids in phase space or from the
density distribution and the available energy range in the model and we measure
the corresponding fraction of different kinds of motion. These percentages can
depend on the choice of the sets of initial conditions and a priori are not
expected to be equal. For this reason, before varying any other parameter we
should consider as carefully as possible what initial conditions to choose and
how well this choice ``spans'' the allowed phase system of the system.

We first considered sets of orbits in such a way as to favor the dynamics near
the main family of periodic orbits x1 (building blocks of the bar). To this
end, we launched initial conditions along the bar's major $x$--axis with
positive momenta $p_y$ and position ($z$) or momentum values ($p_z$) in the
$z$--direction (distributions $I$ and $II$ below). We also considered initial
conditions which follow the model's total density $\rho$ in positions $(x,y,z)$
while their positions are arbitrary (distribution $III$ below).
\begin{table}\label{table:1}
 \begin{center}
 \begin{minipage}{140mm}
 \caption{Varying parameters for all sets of initial conditions.}
 \hspace{0.5cm}
\begin{tabular}{|c|c|c|c|c|c|c|r|r|r|r||r|r|r|r|}
  \hline
  \hline
  distribution/model   & $M_{B}$ & $b$ & $c$\\ \hline
  $IS,IIS,IIIS$ & 0.1 & 1.5 & 0.6\\ \hline
  $IC,IIC,IIIC$ & 0.1 & 1.5 & 1.2\\ \hline
  $IB,IIB,IIIB$ & 0.1 & 3.0 & 0.6\\ \hline
  $IM,IIM,IIIM$ & 0.2 & 1.5 & 0.6\\ \hline
\end{tabular}
\end{minipage}
\end{center}
\end{table}
More specifically, the three different classes (distributions) of initial
conditions we use here are:
\begin{itemize}
    \item distribution $I$:\ 50,000 orbits equally spaced in the space
        $(x,z,p_{y})$ with $x\in [0.0,7.0]$, $z\in [0.0,1.5]$, $p_y \in
        [0.0,0.45]$ and $(y,p_{x},p_{z})=(0,0,0)$.\
\end{itemize}
\begin{itemize}
    \item distribution $II$:\ 50,000 orbits equally spaced in the space
        $(x,p_{y},p_{z})$ with $x\in [0.0,7.0]$, $p_y\in [0.0,0.35]$, $p_z
        \in [0.0,0.35]$ and $(y,z,p_{x})=(0,0,0)$.\
\end{itemize}
\begin{itemize}
    \item distribution $III$:\ 50,000 orbits whose spatial coordinates are
        chosen randomly from the density distribution of our model,
        using the
        \textit{rejection method} \citep{NuRe} within the rectangular box
        $-a\leq x \leq a$, $-b\leq y \leq b$, $-c\leq z \leq c$. We then
        draw a random value of the Hamiltonian $H$ (total energy) in the
        range [0, $H_{max}$], where $H_{max}$ is 1.1 times the escape
        energy $H_{esc}=-0.20$\footnote[1]{The reason we choose an extended
        range of energies is because we wish to study some orbits that are
        in this range but don't necessarily escape during our adopted
        integration time.} defined by the zero velocity curve going through
        the $L_1$ or $L_2$ Lagrangian point. Then we set
        $(p_{y},p_{z})=(0,0)$ and we calculate the $p_{x}$--momenta by the
        relation: $p_{x}=H(x,y,z,p_{y},p_{z})$ (keeping the positive
        solution). Note that in this case instead of giving $p_y$--momenta
        we tried $p_x$. This choice allows us to check how a different way
        of populating the phase space may affect the relative fraction
        of regular and chaotic motion. This third distributions has a better
representation of the density, but still is arbitrary with respect to the
velocities.
\end{itemize}

Details of our distributions/models can be found in Table~1. In the
next sections we start by studying the distribution of regular and chaotic
motion in phase space, for different choices of initial conditions,
different bar parameters (sizes, mass and pattern speed). We also
briefly study the distribution as a function of energy and of location
in configuration space.
\begin{figure*}\centering
\epsfig{file=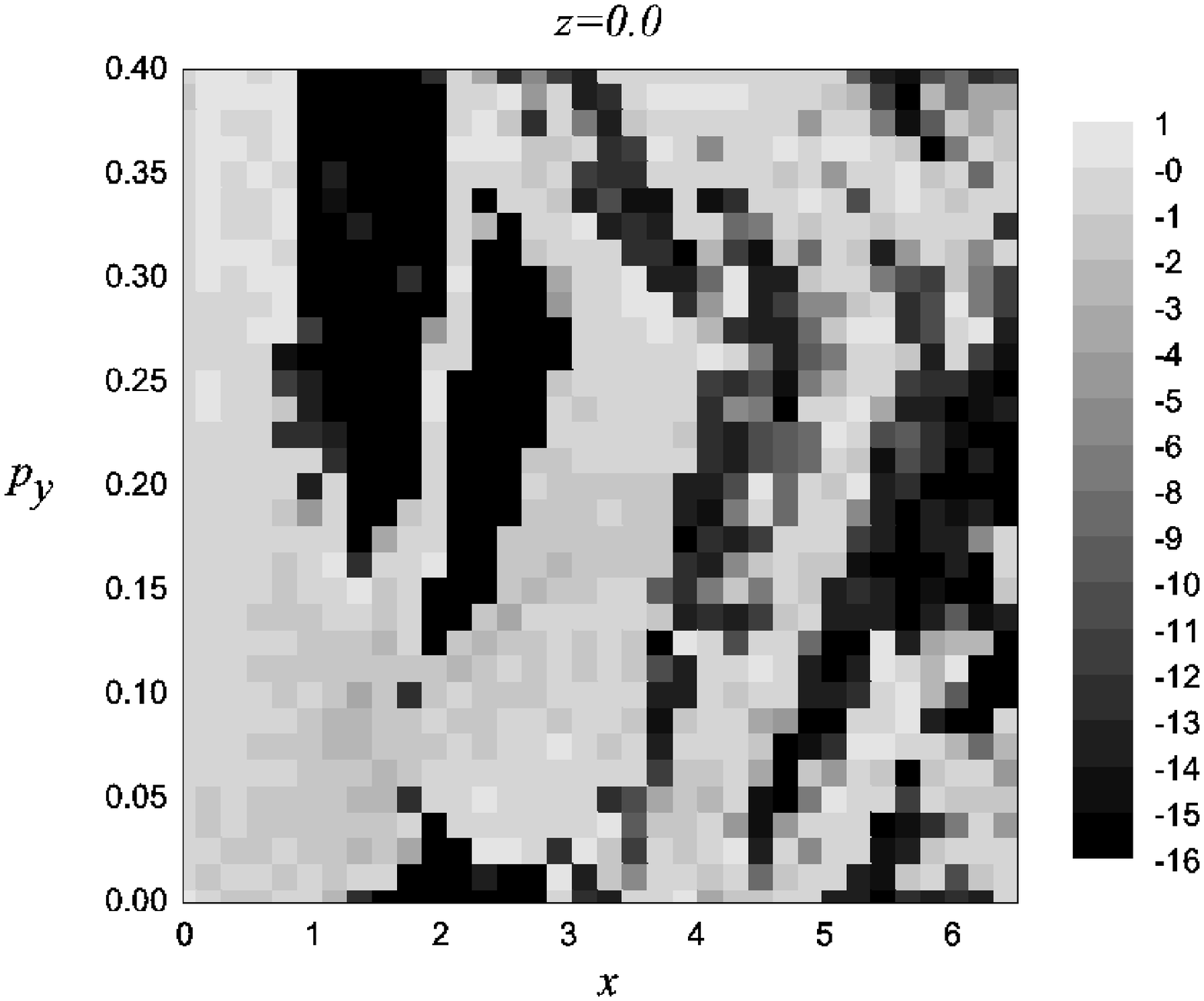,height=4.5cm,width=6.5cm}\hspace{0.5cm}
\epsfig{file=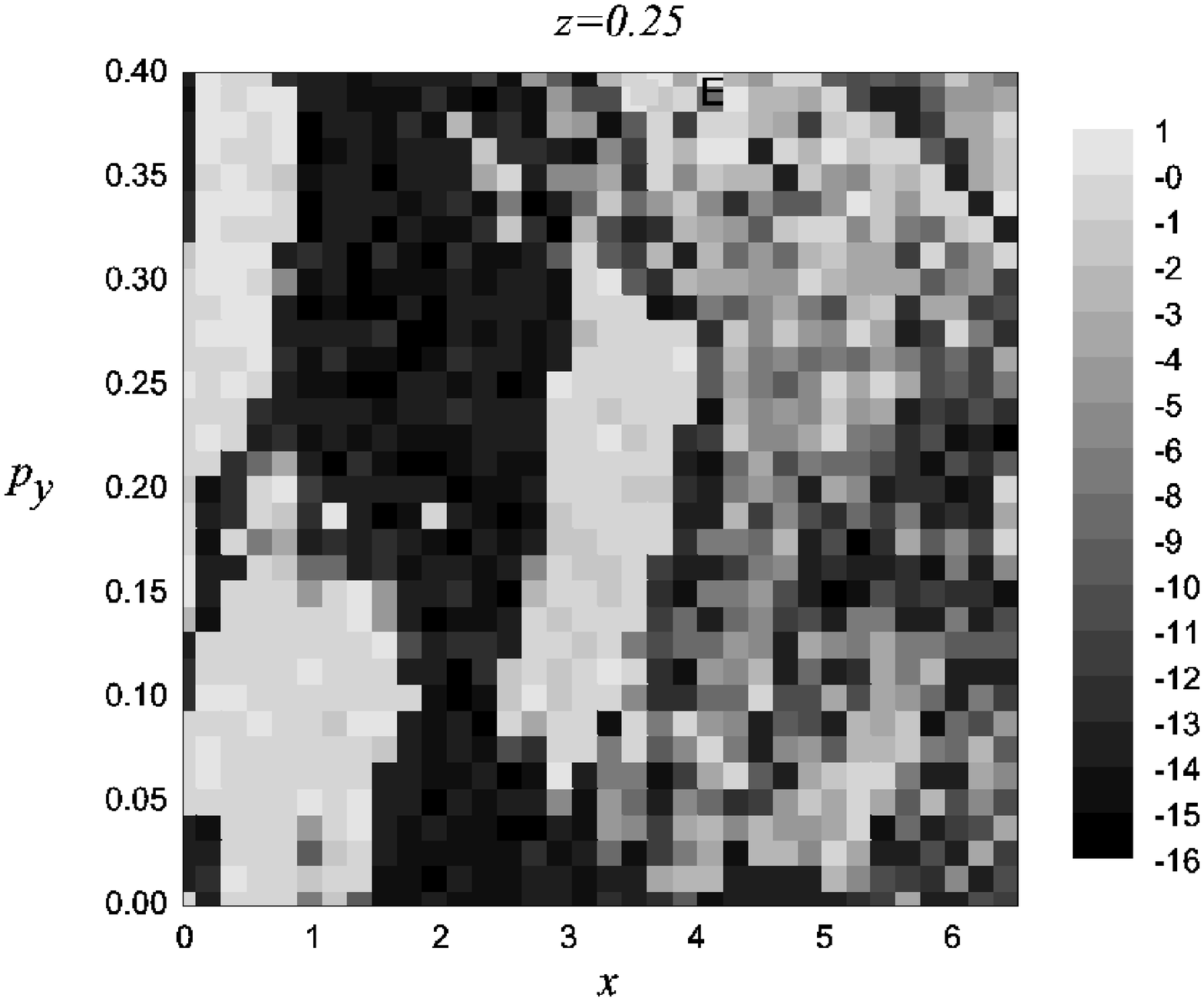,height=4.5cm,width=6.5cm}\\\vspace{0.25cm}
\epsfig{file=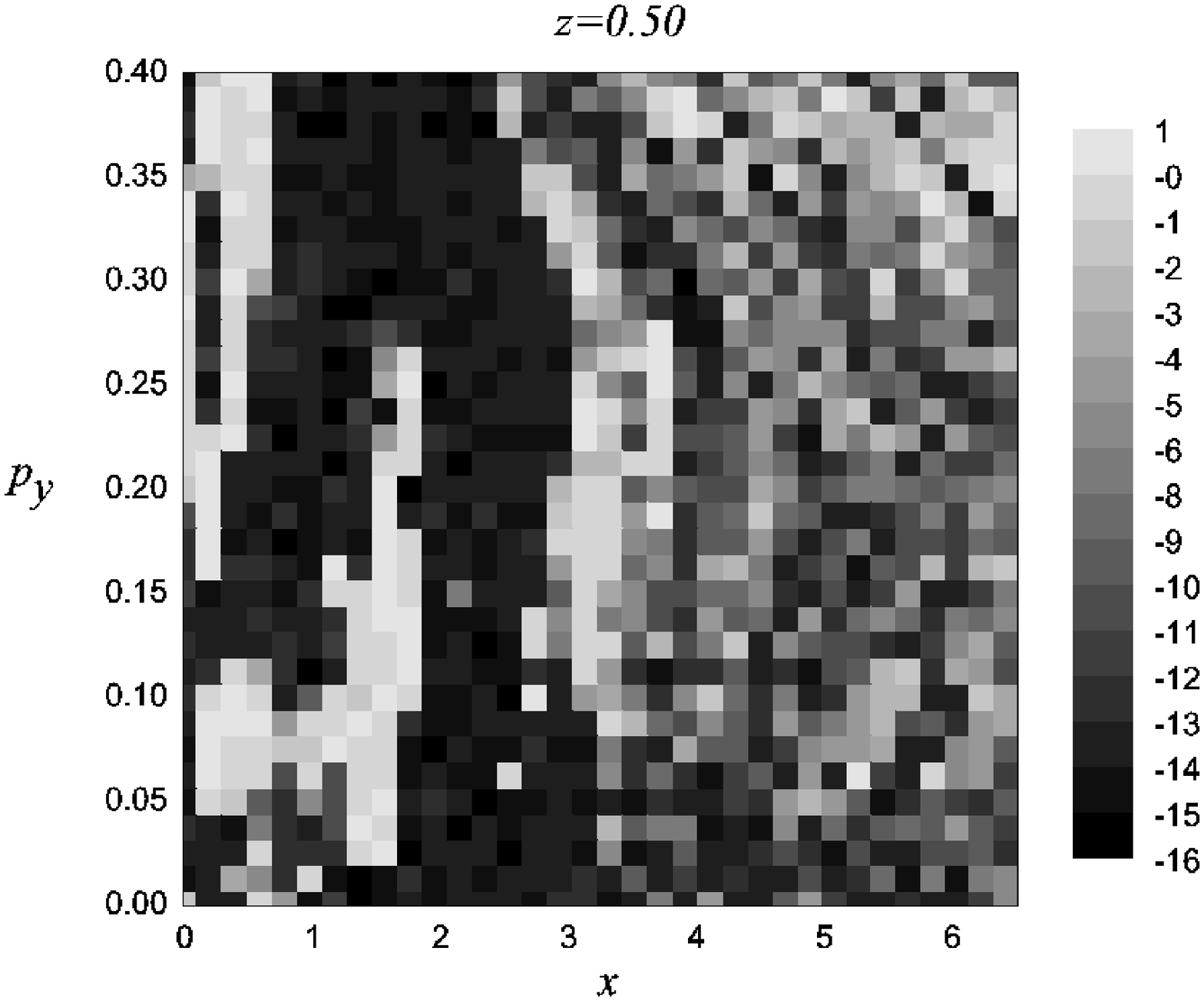,height=4.5cm,width=6.5cm}\hspace{0.5cm}
\epsfig{file=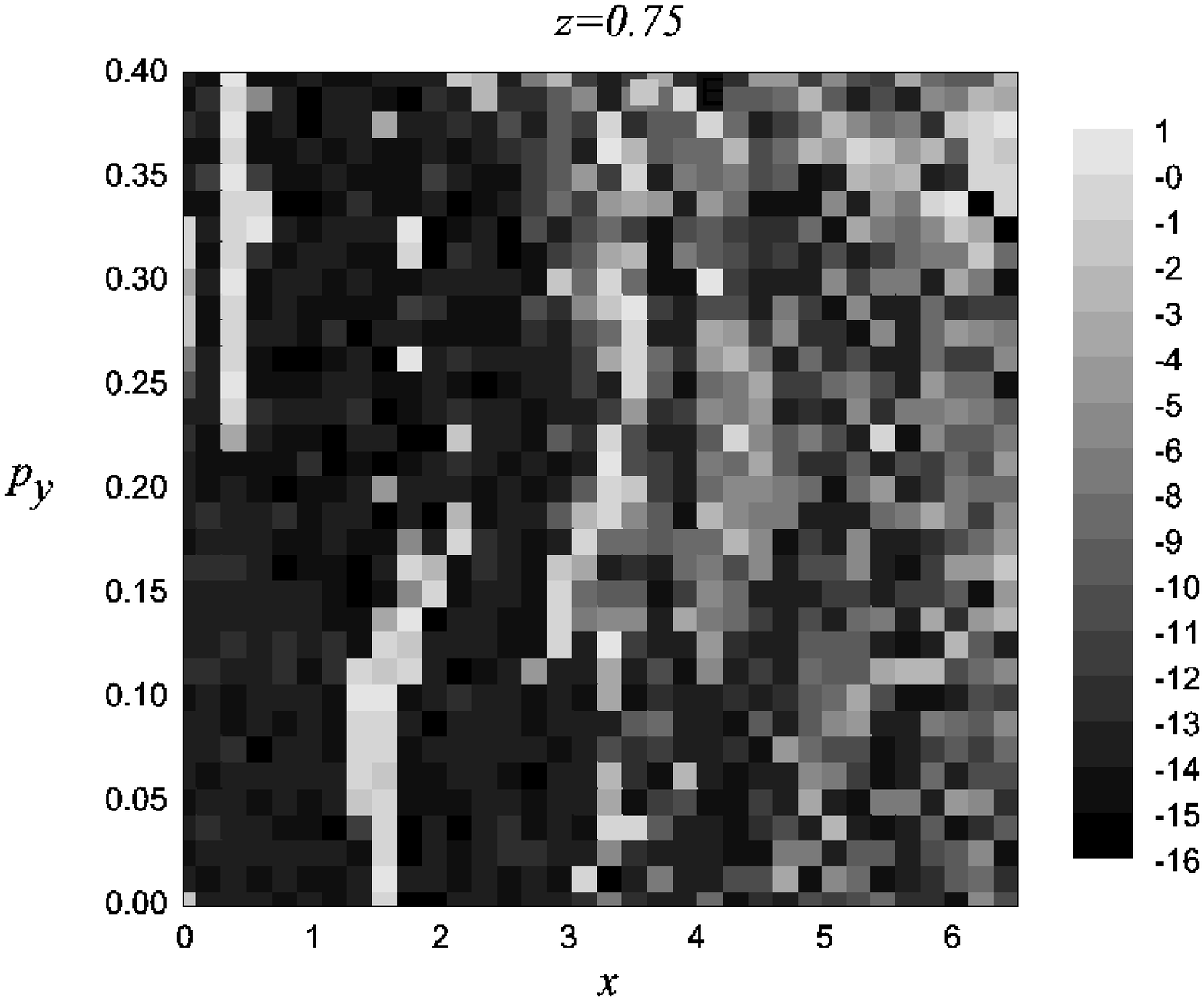,height=4.5cm,width=6.5cm}\\\vspace{0.25cm}
\epsfig{file=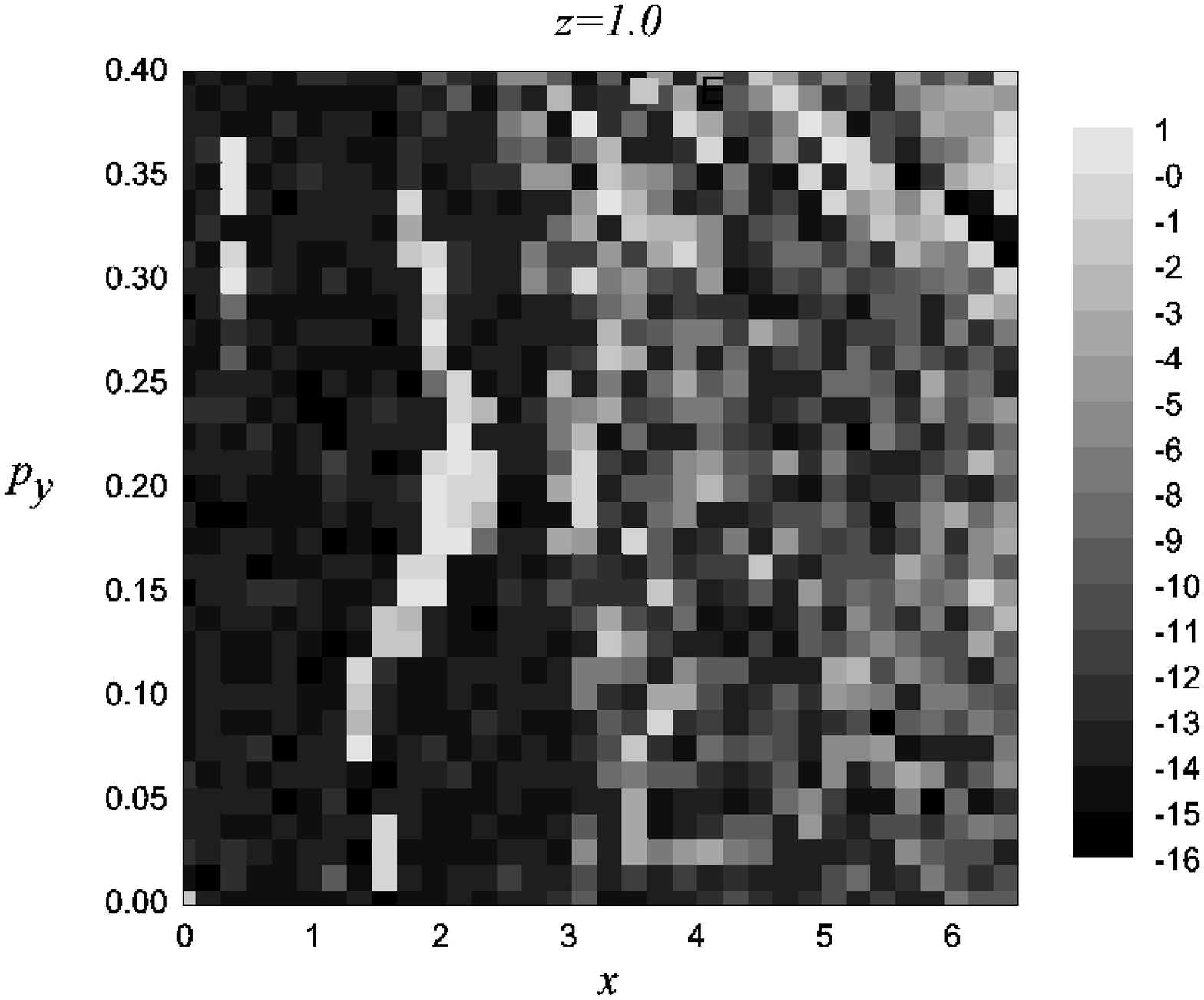,height=4.5cm,width=6.5cm}\hspace{0.5cm}
\epsfig{file=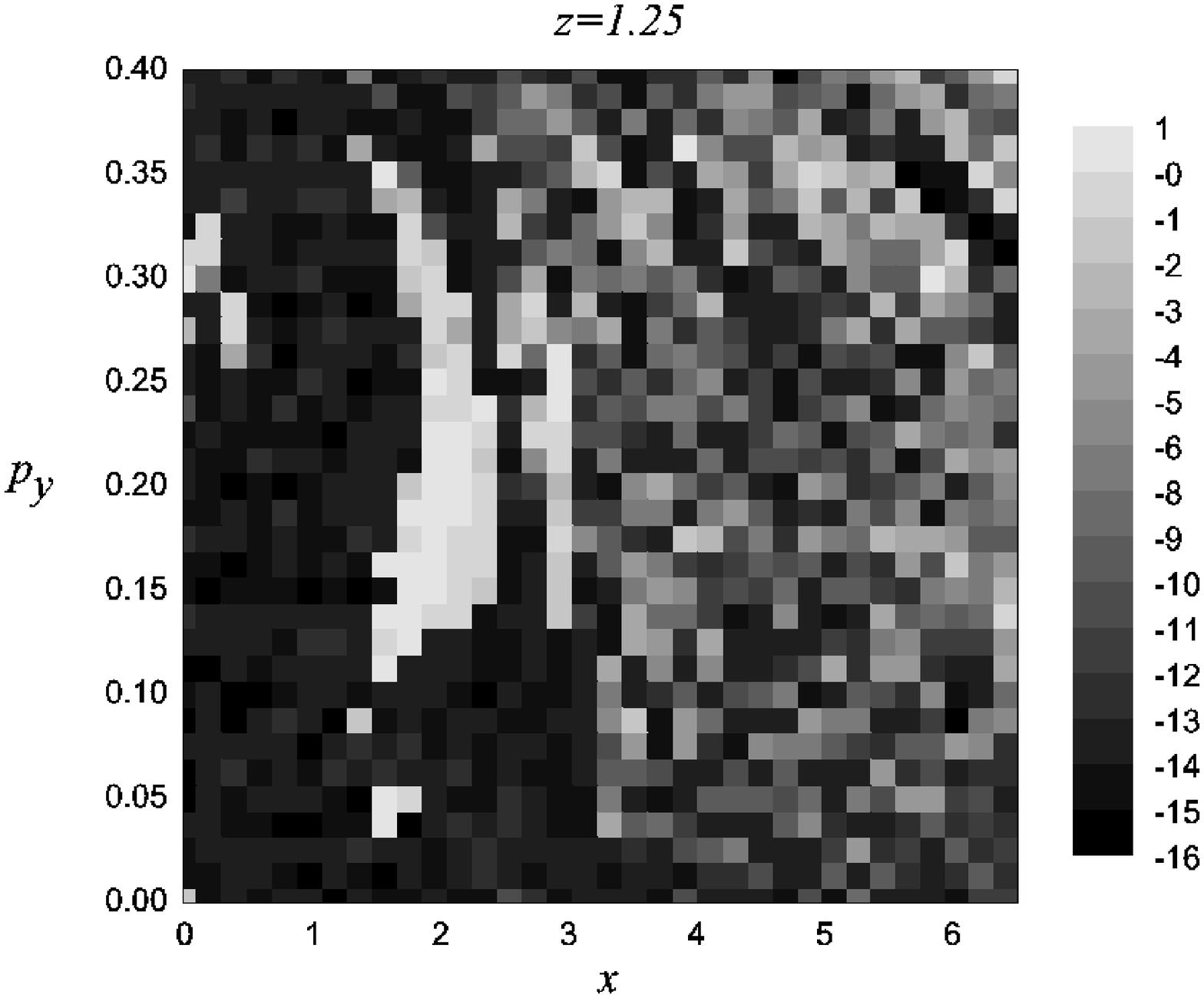,height=4.5cm,width=6.5cm}
 \caption{Slices of the $(x,p_y)$--plane for different $z$ values. Dark regions correspond
to chaotic initial conditions, light gray to regular ones and intermediate colors to orbits 
with a small rate of local exponential divergence, and/or orbits whose chaotic nature is 
revealed after long times, like for example the so-called sticky orbits. The gray-scale bar 
represents the values of the SALI in logarithmic scale.}
\label{percplane}
\end{figure*}

\subsection{Percentages of regular orbits vs. energy}
\label{3dofpercenerg}

We next examine the variation of the percentages of regular and chaotic orbits
as a function of the energy and we show the results in Fig.~\ref{perc_sev}a. We
first divided the energy interval into 30 subintervals, each containing an
equal number of orbits. In every subinterval we calculated the percentages of
regular and chaotic orbits. In Fig.~\ref{perc_sev}a we plot the percentage of
regular orbits for energies up to $H \simeq -0.15$, i.e. up to a value somewhat
beyond the escape energy for all models under study. Generally, we observe that
the fraction of regular orbits decreases as the energy increases. An important
peak, however, occurs before the escape energy, so that for ($H>-0.20$) the
fraction of regular orbits increases again! This non--monotonic behavior is
related to the growth of the islands around some basic stable periodic orbits
in the phase space and is similar to what was observed in the two DOF model,
comparing the fraction of regular orbits for the common energy interval, i.e.
up to $H \leq -0.24$ (see Fig.~\ref{per_EN_2dof}).

Picking another set of initial conditions we find similar qualitative results
with some small quantitative differences. Hence, we find the same behavior when
studying the models of distribution $II$, i.e. regular motion is
dominant for small generally energies, then a decrease and then a peak.
Distribution $III$ manifests the same trend for small energies while the peak
is less intense compared to the other two distributions.

\subsection{Spread of regular orbits in configuration space}
\label{spreadrc}

We also explored the way in which regular and chaotic orbits are distributed
along the $z$--direction of the configuration space. Following the evolution of
each orbit, we calculated the \textit{mean absolute value} of their
$z$--coordinate ($<|z|>$) and plot in Fig.~\ref{perc_sev}b the fraction of
regular orbits as a function of the $<|z|>$ for the models of distribution $I$.
It clearly follows from these results that the intervals relatively near the
$(x,y)$--plane ($<|z|>~< 0.35$) contain mainly regular orbits, while those with
larger values of mean absolute distance $z$ are mostly chaotic. Similar
qualitative results are obtained for distribution $II$ (not shown here).
Checking distribution $III$ we find good match for the relative fraction of
regular motion for relatively small $<|z|>$ (up to 0.4) with the other models.
For intermediate values ($0.4<~<|z|>~<1.5$) the motion is mostly chaotic while
we don't have orbits reaching larger values, which is due to the way the
initial conditions are given.

We also looked at these percentages as a function of the initial
spherical radius ($R_{spherical}=\sqrt{x_0^2+y_0^2+z_0^2}$) and the mean
spherical radius over the evolution of the orbits ($<R_{spherical}>$). Again,
dividing the total range of the $R_{spherical}$ values in 30 subintervals, we
calculated the percentages of regular orbits as a function of the mean
$R_{spherical}$ value in each subinterval. We find that the fraction of regular
orbits for all models decreases sharply with increasing $R_{spherical}$ up to
$R_{spherical}<1.5$, where it reaches a minimum (see Fig.~\ref{perc_sev}c). For
$1.5<R_{spherical}<7$ it starts to increase gradually almost monotonically for
model $IS$ and with some rather week fluctuations for $IC$. Note that the
model $IB$ possesses a relatively large fraction of regular motion within the
interval $2.5<R_{spherical}<4.5$ as well. Model $IM$ follows a similar trend,
but with a different scale than $IB$, containing its larger fraction of regular
motion in the same roughly intervals, while for larger $R_{spherical}$ the motion
is mainly chaotic. This result is in good agreement with the results in
Fig.~\ref{perc_sev}d, where the horizontal axis corresponds to the value of the
mean spherical radius over the evolution time of the orbits. As previously,
distribution $II$ confirms well this behavior while distribution $III$ is in
good agreement for relatively small radial distances from the center of the
galaxy and diversifies for larger ones.

\subsection{Fraction of regular motion as a function of distance from
the equatorial plane}
\label{3dofperczdirection}

We then examine whether the orbital motion is more ``regular'' near the
$(x,y)$--plane or far from it and how this relates to their behavior as a
function of $<|z|>$ discussed in the subsection~\ref{spreadrc} and
Fig.~\ref{perc_sev}b. For this, we took the set of initial conditions from
distribution $I$ (version $IC$ in this example) with initial conditions given
in the space $(x,z,p_y)$ and create a mesh in the $(x,p_y)$--plane with
different slices in the $z$--direction that start from $z=0$ up to $z=1.25$
with $step=0.25$. In Fig.~\ref{percplane}, we see that as the distance of the
initial conditions from the equatorial plane increases (from top to bottom row
and from left to right panels), large islands of stability in the
$(x,p_y)$--plane start to shrink, when $x$ is between 0 and 2. This trend is
almost monotonic up to the $z$--slice $z \approx 1$ (Fig.~\ref{percplane},
third row, left panel). For $z=1.25$ (Fig.~\ref{percplane}, third row, right
panel) the stability region, lying between $x=1.5$ and $x=2$ starts to grow.
This result turns out to be in good agreement with the one in
Fig.~\ref{perc_sev}b, where we plot the percentages of regular orbits as a
function of $<|z|>$. This trend, as expected, is similar for the other two
different sets of initial conditions (distributions $IIC$ and $IIIC$), since
the model parameters remain the same.

\subsection{Percentages of regular orbits vs pattern speed}
\label{3dofpercepattern}

Furthermore, choosing one class of initial conditions (distribution $IS$), we
compare different models for several values of the bar's pattern speed
$\Omega_{b}$, keeping the remaining parameters constant. In this study, we try
to explore the effect of the value of the pattern speed on the percentages of
globally ordered or chaotic behavior of the system.

Based on the structure and crowding of the periodic orbits, \cite{Con:1} showed
that bars have to end before corotation, i.e. that $r_{L}>a$, where $r_{L}$ the
Lagrangian, or corotation, radius. A similar result was reached from the
calculation of the self-consistent response to a bar forcing \citep{Atha80}.
These two results are very useful, since they set a lower limit to the
corotation radius, which can thus not be smaller than the bar length, but give
no information on an upper limit. Comparing the shape of the observed dust
lanes along the leading edges of bars to that of the shock loci in hydrodynamic
simulations of gas flow in barred galaxy potentials, \cite{Athan:2,Athan:3} was
able to set both a lower and an upper limit to the corotation radius, namely
$r_{L}=(1.2 \pm 0.2)a$. This restricts the range of possible values of the
pattern speed for our model, from $\Omega_{b}=0.0367$, that corresponds to the
Lagrangian radius $r_{L}=1.4a$, to $\Omega_{b}=0.0554$, that corresponds to
$r_{L}=1.0a$. Using these extreme values, and the three intermediate
frequencies: $\Omega_{b}=0.0403$, $\Omega_{b}=0.0444$ and $\Omega_{b}=0.0494$,
that correspond to the Lagrangian radii $r_{L}=1.3a$, $r_{L}=1.2a$ and
$r_{L}=1.1a$, respectively, we investigated how the value of the pattern speed
affects the system and found that as $\Omega_{b}$ increases the percentage of
the regular orbits decreases relatively weakly. The corresponding results are
given in Table~2. It turns out that the variation of the pattern speed does not
affect drastically the relative fraction of regular and chaotic motion. One
would probably need to try more extreme values of $\Omega_b$, beyond the upper
and lower limits of the corotation radius, in order to see a significant
change. Such values, however, would be unrealistic for real galaxies.
\begin{table}\label{table:2}
\begin{center}
\begin{minipage}{140mm}
\caption{Models with different pattern speeds $\Omega_{b}$.} \hspace{0.15cm}
\begin{tabular}{|c|c|c|c|l|l|l|l|l|l|l|l|l|l|l|}
  \hline
  \hline
  model          & $\Omega_{b}$ & $r_{L}$ & \% of Regular orbits \\ \hline
  $I\Omega_{1}$  & 0.0367032 & 1.4a & 28.39 \\       \hline
  $I\Omega_{2}$  & 0.0403014 & 1.3a & 27.63 \\       \hline
  $I\Omega_{2}$  & 0.0444365 & 1.2a & 27.26 \\       \hline
  $I\Omega_{4}$  & 0.0493654 & 1.1a & 26.55 \\       \hline
  $I\Omega_{5}$  & 0.0554349 & 1.0a & 25.55 \\       \hline
\end{tabular}
\end{minipage}
\end{center}
\end{table}
\subsection{Fraction of regular and chaotic trajectories}
\label{3dofpercfraction}

In this section and in Fig.~\ref{ABC}, we investigate the amount of regular
motion in the phase space for all the three sets of initial conditions
(distributions $I$, $II$ and $III$), varying the axial ratios ($b/a$ or $c/a$
parameters) and the mass ($M_B$ parameter) of the bar component, as described
in Table~1. In all distributions we used 50,000 initial conditions and we
employed the \textit{bootstrap}--method \citep{NuRe} on several subsamples to
make sure that this number is sufficiently high to provide reliable
information. Computing the distributions of regular and chaotic orbits for the
above classes of initial conditions we find the following:

\subsubsection{Percentages for distribution $I$}

Before starting with the full 3D model, we first measured the percentages of
regular orbits in the 2D model. We use initial conditions equally spaced in
$(x,p_{y})$, with $(y,p_{x})=(0,0)$, $x\in [0.0,7.0]$ and $p_y \in [0.0,0.45]$.
We find relative fractions of regular motion equal to $45.40\%$ for model $IM$,
$53.43\%$, for model $IS$, $92.43\%$ for model $IB$ and $64.43\%$ for model
$IC$. This shows that at least in the 2D case, the massive bar model ($IM$) has
the highest fraction of chaotic orbits, followed by the standard distribution
$IS$. The two models with the extended short ($IC$), or intermediate axis
($IB$) have the least chaos.

\begin{figure*}
{\includegraphics[height=0.225\textheight]{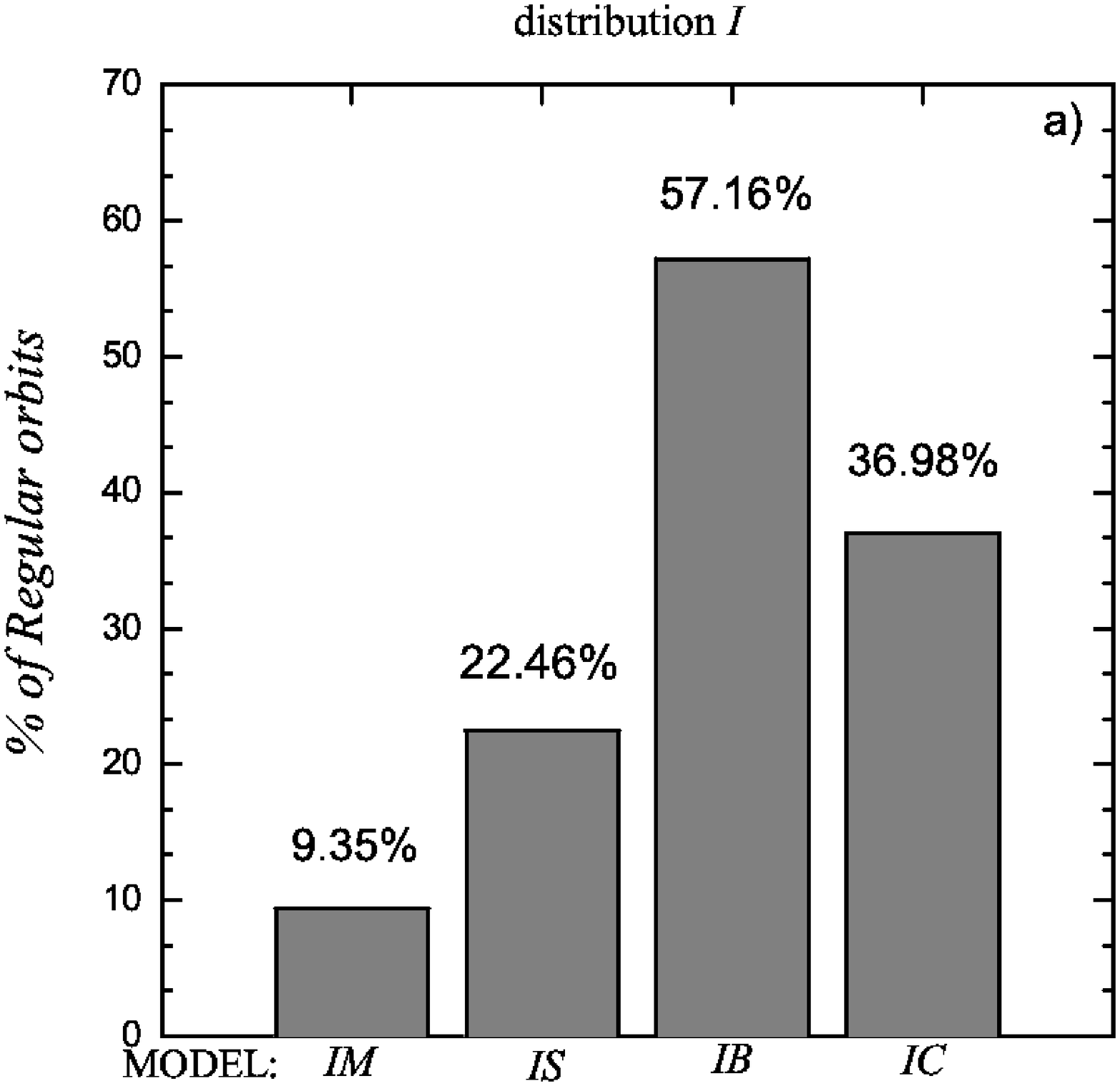}}
{\includegraphics[height=0.225\textheight]{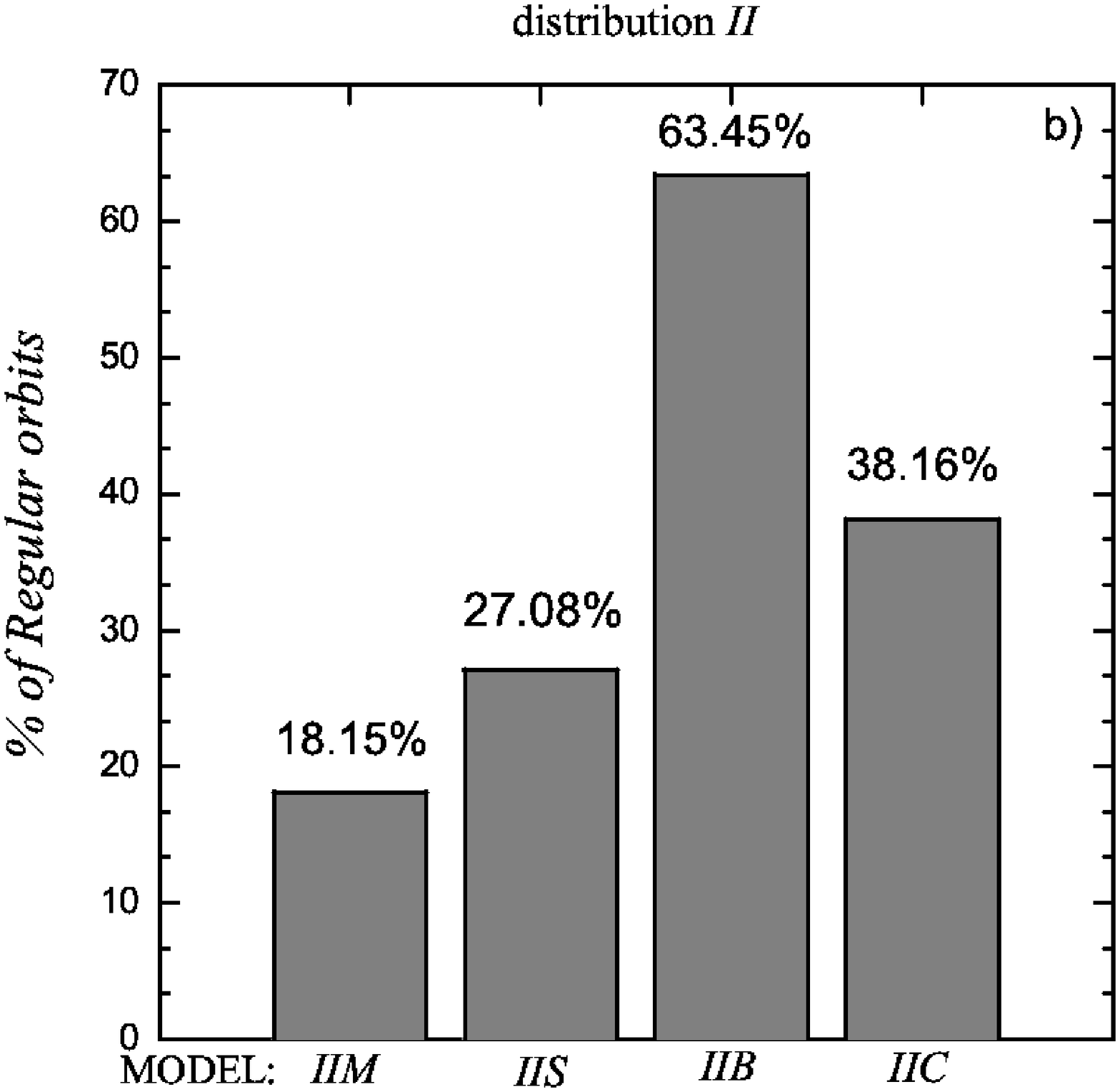}}
{\includegraphics[height=0.225\textheight]{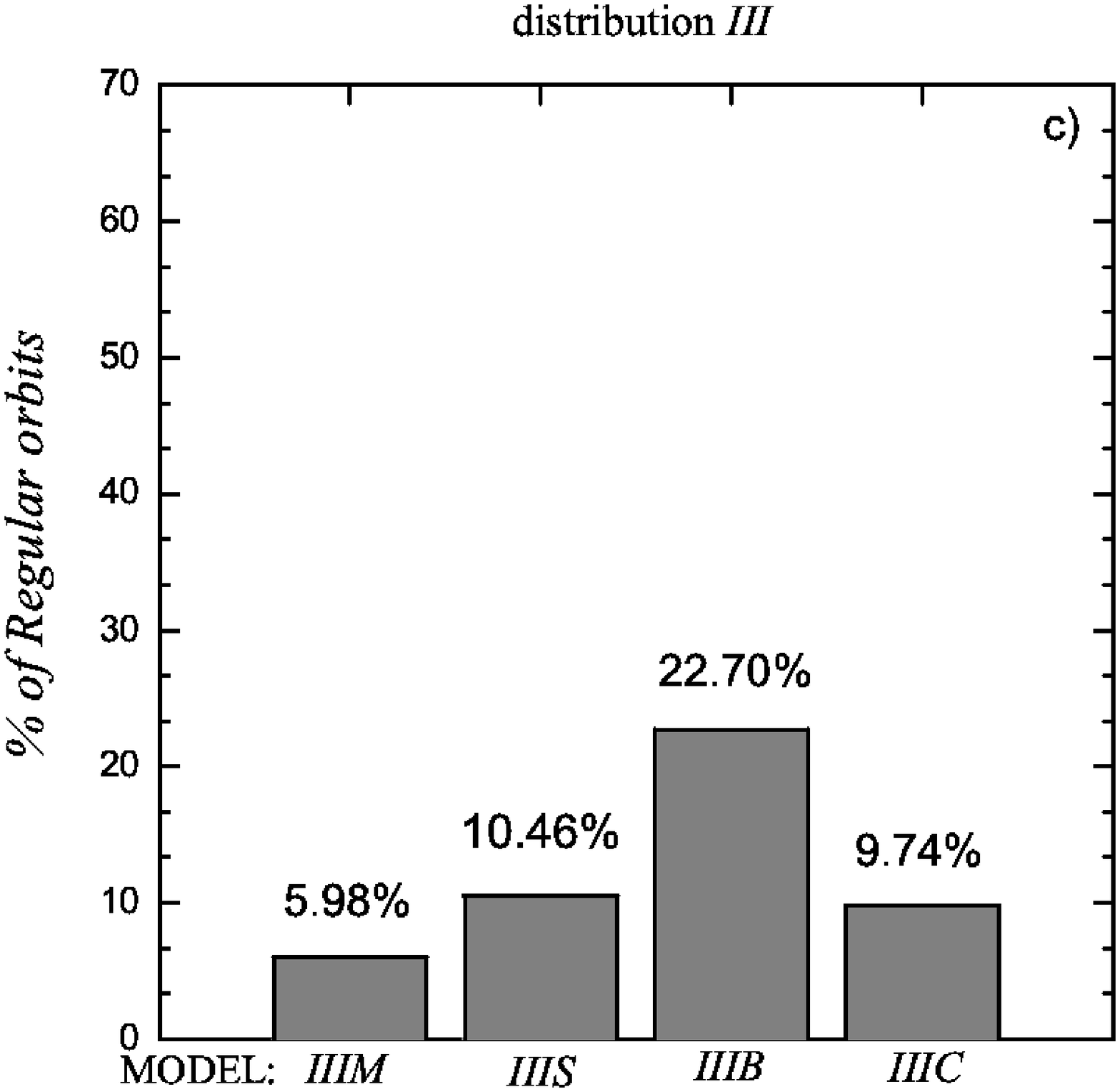}}
\caption{Percentages of regular orbits (SALI$\geq10^{-8}$) for distribution $I$
(left panel), $II$ (middle panel) and $III$ (right panel) of initial
conditions for all the different model versions, where the bar mass and the
length of its intermediate and minor (vertical) semi-axes are varied. Generally,
when the bar mass increases the model tends to be more chaotic, while when the
size of the bar semi-axes ($b$ or $c$ parameters) increases the fraction of
regular motion grows. The small difference in the trends between models $IS
\rightarrow IC$ and $IIIS \rightarrow IIIC$ is due to the different way in
which the sample of initial conditions in phase space is chosen.} \label{ABC}
\end{figure*}

Let us now turn to the full 3D orbital coverage. In Fig.~\ref{ABC}a we present
the percentages of the regular orbits for the various sets of initial
conditions of the distribution $I$. Comparing the percentages, we see that
increasing the mass of the bar in model $IM$ increases chaotic behavior. This
confirms the results obtained above and in \cite{ABMP} for the two DOF case. On the other
hand, when the bar is thicker (model $IC$), i.e. the length of the $z$--axis
larger, the system becomes more regular. Finally, the corresponding results for
a fatter bar, i.e. with larger $y$--axis (model $IB$) demonstrate that the
increase of the intermediate axis of the bar also provides the system with more
ordered behavior.
Comparing the 2D and 3D cases presented above, we note that the trends we find
are the same, but that the relative fractions of regular orbits are somewhat
higher in the 2D case. We will discuss this further towards the end of the
section.

\subsubsection{Percentages for distribution $II$}
Similarly, in Fig.~\ref{ABC}b, we present percentages of the regular orbits for
initial conditions selected for distribution $II$ and models $S$, $M$, $B$ and
$C$. As before, the increase of the bar mass (model $IIM$) causes more
extensive chaos, while for a thicker bar or fatter bar (models $IIB$ and
$IIC$), the system is again more regular. It is thus clear that the results of
distributions $I$ and $II$ are similar, presumably dues to the fact that they
contain initial conditions that cover the phase space in a similar manner, i.e.
they favor the region around the x1 family of periodic orbits.

\subsubsection{Percentages for distribution $III$}
Finally, we plot in Fig.~\ref{ABC}c the percentages of regular orbits for
several sets of initial conditions from distribution $III$. A first general
observation is that the fraction of ordered motion is significantly smaller
than is distributions I and II, for all models. The basic reason for this
difference is related to the way that the momenta of the positions are given in
this distribution of initial conditions. In particular, all the orbits in this
case are launched with a $p_x$ momentum instead of a $p_y$ as in distributions
$I$ and $II$. As a consequence, the big island of stability around the main
family of periodic orbits x1 is populate with fewer orbits and thus in general
more chaotic orbits are launched and measured. Nevertheless, again the increase
of the bar mass in the $IIIM$ model results in more chaotic behavior. As for
the $IB$ model, a thicker bar in the $y$--direction turns out to make the
system more regular. Regarding the case of larger $z$--semiaxis (model $IIIC$),
we notice a small decrease in the percentages of regular orbits (9.74$\%$ now,
from 10.46$\%$). It turns out that this particular different distribution of
initial conditions doesn't reveal quite the same trend as in the previous cases
($IS \rightarrow IC$ and $IIS \rightarrow IIC$).

Concluding this section dedicated to the dynamical study of regular and chaotic
motion in the phase space is that their corresponding fraction is basically
related to the way that one populates the main stable (or unstable) periodic
orbits of the system. We find that motion near the plane is generally stable
(non chaotic) and that trajectories spending the largest fraction of time at
large $z$ are usually chaotic (with large Lyapunov exponents). These results
are in agreement with the results obtained by \cite{ZS1,ZS2} for different
models. The increase of the bar mass leads to more chaotic motion in the phase
space as found in \cite{ABMP} for a two DOF model, while in general the
increase of the bar $y$ or $z$--semiaxis leads to more regularity.
\begin{figure}\centering
  \includegraphics[height=0.275\textheight]{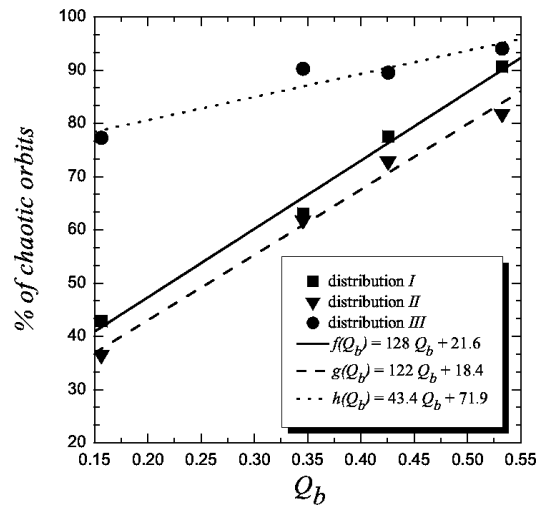}
  \caption{The relative fraction of chaotic motion as a function of bar
  strength $Q_b$ increases (see Fig.\ref{barstre}). Filled squares
  ($\blacksquare$) correspond to the models $M$, $S$, $C$ and $B$
  (from right to left, respectively) for distribution $I$. while filled triangles
  (\large{$\blacktriangledown$}\small) are for distribution $II$ and filled circles
  (\Large $\bullet$\small) for distribution $III$.}
  \label{Qbvschaotic}
\end{figure}
\subsection{The effect of bar strength}
Reviewing the above percentages of regular (and chaotic) motion in phase space,
we see that they change in the same way when the basic parameters and
properties of the model are varied. In order to understand this better we
examine next the effect of the bar strength, measuring it as discussed in
Section~\ref{Model_gal}, i.e. from the relative strength of the
non-axisymmetric forcing. In Fig.~\ref{Qbvschaotic} we plot the relative
fraction of chaotic motion as a function of bar strength $Q_b$. Black filled
squares correspond to distribution $I$, filled triangles to distribution $II$
and filled circles to distribution $III$. This figure reveals that, provided
the choice of initial conditions for the orbits is the same or similar, there
is a tight correlation between the two quantities. the amount of chaos
increases with increasing bar strength, as intuitively expected. What is
unexpected, however, is how tight all these correlations are, with rank
correlation coefficients $\sim -0.936$ for all distributions.

Fig.~\ref{Qbvschaotic} also shows that, except for the bar strength, the
distribution of initial conditions chosen is also important. Distributions $I$
and $II$ differ little and their best fitting straight lines are roughly
parallel and little displaced from each other. This is in good agreement with
our previous statement (subsect.~\ref{inconcases}) that they both cover well
the regular orbits trapped around the stable orbits of the x1 family, which are
in fact the backbones of the bar \citep{ABMP}. On the other hand, distribution
$III$ has a much higher fraction of chaotic orbits, as could also be seen from
Fig.~\ref{ABC}c, attesting that the corresponding distribution of
initial conditions is more chaotic, for the reasons we already described.

\section{Conclusions}
\label{Conclusions}

In this paper, paper I, we studied the detection and distribution of regular and
chaotic motion in the phase space of barred galaxy models. Our results
referring to the dynamical properties of the 2D/3D model with a
Ferrers bar can be
summarized as follows:
\begin{enumerate}
  \item We applied successfully the GALI method to distinguish both
      qualitatively and quantitatively between regular and chaotic orbits.
      We showed its efficiency and its advantage in detecting fast and
      accurately the chaotic nature of a trajectory.
  \item In 2D systems, using the SALI (GALI$_2$) method, we were able to
      identify efficiently and fast tiny regions of regular or chaotic
      motion, which are not clearly visible on PSSs.
  \item Using GALI$_{2,3}$ we detected regular motion in low dimensional
      tori, i.e. examples of regular orbits of the three DOF Ferrers' model
      that lie on a 2D torus, while the torus' expected dimension is
      generally three. Concerning chaotic orbits, the GALI$_{3,4,5,6}$ indices
      decay faster than SALI and are able to detect chaos at early times well
      before this is evident from the calculation of the mLCE.
  \item We also tried different models and different distributions of
      initial conditions for the orbits to test the fraction of regular and
      chaotic motion in various cases. We found that regular orbits are
      generally dominant at relatively small radial distances from the
      center of the galaxy and at small distances from the equatorial
      plane.
  \item We tested the effect of varying the bar pattern speed for our
      standard model and found that, within a realistic range of values,
      this parameter does not affect much the phase space dynamics. Within
      these limits, we found that the fraction of regular motion varies
      only by about 3 per cent, in the sense that the bars at the slow
      limit of the realistic range have more regular motion than the bars
      in the fast limit.
  \item We varied the values of the main parameters of our models, such or
      the mass and the axial ratio of the bar component (in the 3D model). Chaos
      turns out to be dominant in galaxy models whose bar component is more
      massive, while models with a thicker or fatter bar present generally
      more regular behavior although the initial conditions are given can
      in general affect somewhat the relative percentages.
  \item We found a very strong correlation (rank correlation coefficient at
      $\sim$0.94) between the fraction of chaotic orbits and the relative
      strength of the non-axisymmetric forcing. This holds for all our
      three
      initial conditions distributions taken individually, even though they
      were specifically chosen so as to represent different types of
      orbits. We can thus conclude that strong non-axisymmetric forcings
      (i.e. strong bars) are the main cause of the presence of large amount
      of chaotic motion in phase space.\

\end{enumerate}

In our next paper of this series (paper II) we focus on the significance of the
``different degrees" of chaotic motion, like the strong and weak chaos, and
their significance from an astronomical point of view. The fact that many
``sticky"/chaotic orbits often persist and behave in a regular manner for very
large time intervals, before showing their chaoticity, makes them
astronomically important for supporting the galaxy structure. We present there
a first approach in filtering these weak chaotic orbits from the strongly
chaotic ones and classify them observationally as effectively ordered motion.

\section{Acknowledgements}
We would like to thank Ch.~Skokos, T.~Bountis, A.~Bosma, M.~Romero-G\'{o}mez
and P.A.~Patsis for their fruitful comments and discussions on this work. We
also thank the referee, P.M.~Cincotta, for his report, which helped us improve
the clarity of the paper. T.~Manos acknowledges the ``Karatheodory" graduate
student fellowship No B395 of the University of Patras and funds from the Marie
Curie fellowship No HPMT-CT-2001-00338 to Marseille observatory. Visits between
the Marseille and Patras team members were partially funded by the region PACA
in France. This work was partially supported by the grant ANR-06-BLAN-0172.


\end{document}